\documentclass[11pt]{article}

\usepackage{amsmath,amssymb,amsthm}
\usepackage{bm}
\usepackage{algorithm}
\usepackage{algpseudocode}
\usepackage{graphicx}
\usepackage{booktabs}
\usepackage[round]{natbib}

\usepackage[a4paper,margin=25mm]{geometry}
\usepackage{xcolor}
\usepackage[colorlinks=true, allcolors=blue]{hyperref}

\newcommand{\R}{\mathbb{R}}
\newcommand{\Om}{\Omega}
\newcommand{\x}{\bm{x}}

\newcommand{\proj}[1]{\mathcal{P}_{#1}}

\begin{document}

\title{Just-in-time volumetric imaging: Learning 2D-3D thoracic motion models during image-guided interventions}

\author{Nicholas Hindley\\[2pt] {\small Image X Institute, School of Health Sciences, Faculty of Medicine and Health, University of Sydney}\\ {\small \texttt{nicholas.hindley@sydney.edu.au}}}
\date{}

\maketitle

\begin{abstract}
Despite two decades of innovation, cutting-edge bronchoscopic techniques still fail to yield clear diagnostic decisions in $30\%$ of cases and miss malignancies in $20\%$ of cases. Intraprocedural 3D imaging drastically improves both tool positioning and diagnostic yield. However, each 3D spin represents a static snapshot of an anatomy that is continuously evolving as the patient is ventilated throughout the procedure. Here, we propose \emph{just-in-time volumetric imaging}, in which a patient-specific 2D--3D motion model is trained on pre-intervention scans before it is needed and then supplies volumetric images from single fluoroscopic projections during sampling. Given the inspiration-hold CBCT already acquired for navigation and an additional motion-inclusive CBCT, our optimization regime named \emph{VoxelMap-DualScan} (\emph{VM-DS}), solves a low-rank motion model by alternating least squares and then amortizes it into a projection encoder trained on synthesized projections with known motion, all within a $10$-minute window. On eight held-out cases from the public DIR-Lab 4DCT dataset, \emph{VM-DS} achieved a mean landmark target registration error of $2.8 \pm 0.5$~mm over the acquisition arc and $2.8 \pm 0.6$~mm at a fixed anterior--posterior angle, a structural similarity index of $0.95 \pm 0.02$ and a peak signal-to-noise ratio of $37.7 \pm 4.2$~dB. Accuracy was maintained with a $10$-fold reduction in the number of projections required for training from $227$ to $23$. Inference takes $1.5$~ms per projection, making the framework suitable for real-time use during image-guided interventions. Our code is available at https://github.com/Image-X-Institute/Voxelmap/tree/VM-DS
\end{abstract}

\section{Introduction}
\label{sec:introduction}

Lung cancer remains the leading cause of cancer death worldwide, with
approximately $2.5$~million new diagnoses and $1.8$~million deaths per year
\citep{Bray2024}. Outcomes depend overwhelmingly on the stage
at which disease is found, with low-dose CT screening reducing lung cancer mortality in high-risk
populations \citep{NLST2011,deKoning2020}. Screening programmes are
consequently generating a large and growing population of indeterminate
pulmonary lesions that require diagnostic decisions. Guidelines recommend
non-surgical biopsy for indeterminate lesions larger than $8$~mm
\citep{Gould2013,Rivera2013}, and two primary routes are available. Transthoracic
needle aspiration offers high sensitivity for malignancy but at high risk of complication, with pneumothorax occurring in 1 in 4 cases \citep{Huo2020}. Alternatively, bronchoscopy has an excellent safety profile and
permits concurrent mediastinal staging, with fewer procedures, lower
cost and fewer futile surgeries but with lower sensitivity for malignancy \citep{Vakil2021,Ost2020Quality}. Closing the performance gap while retaining the safety and staging advantages of the bronchoscopic route is therefore a central problem in peripheral nodule diagnosis, with improvements in sensitivity for malignancy
translating directly into clinical and economic value
\citep{Ost2024Economic}.

Despite technological advancement over the last two decades, cutting-edge bronchoscopic techniques still fail to yield clear diagnostic decisions in $30\%$ of cases and miss malignancies in $20\%$ of cases. In the TARGET trial, the largest prospective evaluation
of electromagnetic-navigation robotic-assisted bronchoscopy to date, the operator
reached the target lesion in $98.7\%$ of $679$ cases, yet the diagnostic yield
under the standardised ATS/CHEST definition \citep{Gonzalez2024} was only
$61.6\%$ and sensitivity for malignancy was $78.8\%$ \citep{Murgu2025TARGET}.
Similarly, PRECIsE, the multicentre prospective evaluation of shape-sensing robotic-assisted bronchoscopy \citep{Simoff2021,Reisenauer2022}, reported
navigation success of $98.0\%$ but an ATS/ACCP diagnostic yield of $74.1\%$ and
a two-year sensitivity for malignancy of $81.3\%$ across $305$ procedures
\citep{Ost2026PRECIsE}. Non-robotic electromagnetic navigation performed less
favourably, at $62.6\%$ two-year sensitivity in NAVIGATE \citep{Folch2022}.
Across more than ten thousand patients, meta-analyses place pooled yields around
$70\%$ \citep{Kops2023,Nadig2023}.

The deficit in yield and sensitivity versus navigation success can be attributed to tool positioning. Indeed, diagnostic yield is known to fall monotonically with distance from lesion centre \citep{Maturu2025}. Where intraprocedural 3D confirmation is
introduced, tool-in-lesion rates and diagnostic yield both improve drastically.
Tool-in-lesion increased from $34.3\%$ to $98.6\%$ and yield from $29.9\%$ to
$86.6\%$ once mobile cone-beam CT (CBCT) imaging was introduced
\citep{Bashour2024}. Additionally, Husta \emph{et al.} found that $90.7\%$ of initially unsuccessful placements were
converted to tool-in-lesion following repeated spins \citep{Husta2024}. Notably, shape-sensing robotic bronchoscopy alone placed the
needle at the centre of the target in only $26.5\%$ of $102$ nodules, at the
edge or outside in $55.9\%$, with CBCT-informed repositioning recovering a mean
of $4$~mm toward the centre per adjustment \citep{Revelo2025}.

CBCT has become the reference solution in high-volume centres \citep{Setser2020,Bruinen2025}.
However, each spin is a static snapshot of an anatomy that is continuously evolving as the patient is ventilated throughout the procedure. Contemporary series report an average of two spins, either per procedure including an initial planning spin \citep{Maturu2025}, or per lesion to centre the tool \citep{Revelo2025}, so the operator receives two
instants of volumetric information in a procedure lasting $30$--$80$~minutes
\citep{Maturu2025,Chambers2023}. Each
spin also carries dose, table time and, for fixed systems, infrastructure that is only available to a minority of centres \citep{Bashour2024}. Augmented fluoroscopy and digital tomosynthesis provide
additional information by overlaying a segmented target onto live 2D fluoroscopic
images \citep{Aboudara2020,Setser2020}, but the overlay remains static.

Deep learning has recently shown promise in providing real-time volumetric imaging; however, these techniques have been developed
exclusively within the field of radiotherapy, where training data from treatment planning are acquired days or weeks prior to treatment delivery. Patient-specific networks can recover a CT
volume from fixed-angle projections \citep{Shen2019}, subspace and deformation-field
formulations achieve sub-millimetre real-time tumour localisation from single
X-ray projections \citep{Li2010}, and generative and mesh-based approaches
reconstruct volumes or organ surfaces on the fly
\citep{Lei2020TransNet,Wang2019DeepOrganNet}. We proposed the \emph{VoxelMap} deep learning framework that achieves image-modality agnostic volumetric imaging but requires pre-treatment 4D training data \citep{hindley2023,hindley2025}. Subsequently, Shao \emph{et al.} proposed \emph{DREME} \citep{shao2025dreme}, which jointly learns 3D image reconstruction and motion tracking from a single CBCT, but requires a complex 7-stage training curriculum. These existing techniques require between $4$--$20$~hours of training.

\begin{figure}
    \centering
    \includegraphics[width=1\textwidth]{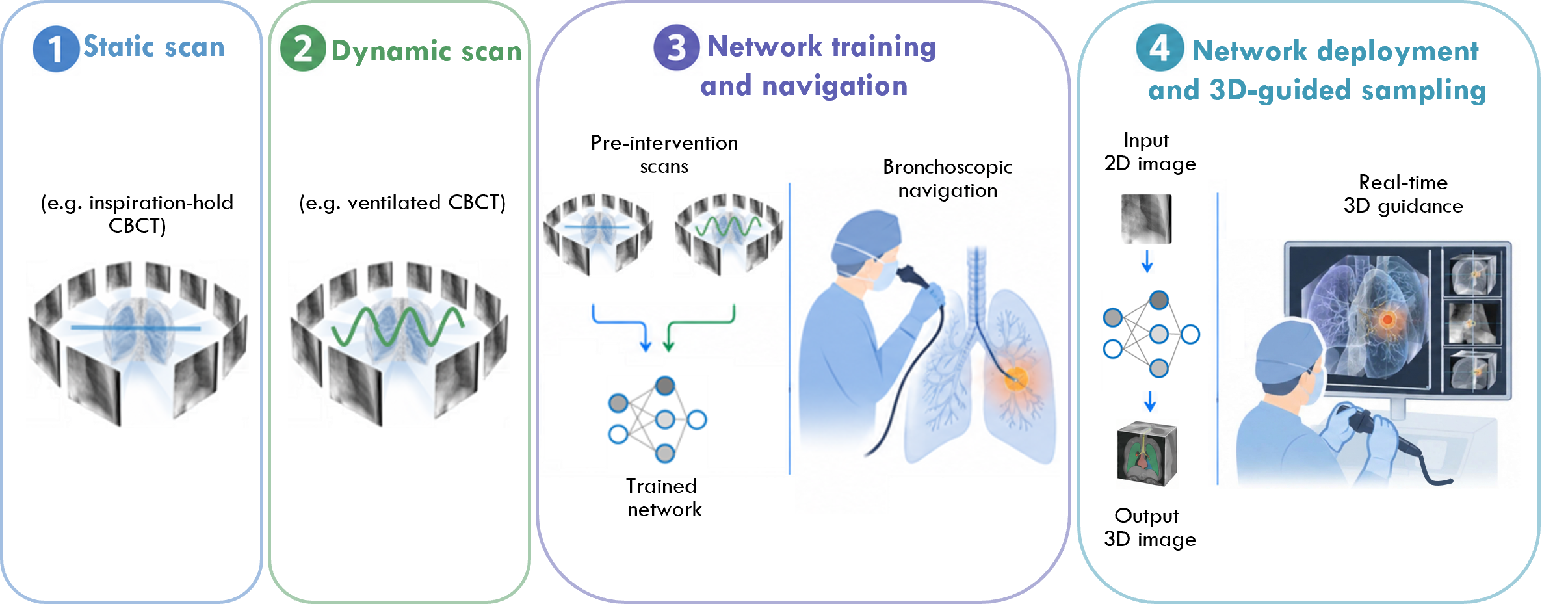}
    \caption{Our proposed clinical workflow. 1. An inspiration-hold CBCT or single phase CT supplies a reference 3D image and is required in standard workflows. 2. A ventilated or free-breathing CBCT supplies x-ray projections from which patient-specific motion can be learned. 3. The network is trained on the pre-intervention scans, completing before sampling begins. 4. The trained network is deployed to provide instantaneous 3D imaging during sampling}
    \label{fig:fig1}
\end{figure}

Training a patient-specific model between the pre-intervention scans and sampling imposes a constraint that is absent from previous work on volumetric imaging. We propose that the training time budget is the interval between the completion of 3D image acquisition and successful navigation to the lesion. Phase-resolved timing places navigation at a median of
$4$~minutes (IQR $3$--$6$), the first CBCT spin at $5$~minutes (IQR $4$--$7$)
and the first needle pass at a further $4$~minutes (IQR $2$--$5$) in a
robotic workflow \citep{Revelo2025}, with conventional CBCT-guided
bronchoscopy slower still \citep{Maturu2025}. If both scans are acquired
before navigation, the model is needed only at the first needle pass, after
navigation and the confirmation spin. These results suggest a window
of approximately $5$--$10$ minutes, and longer windows are of
limited value, as the reference itself loses relevance while atelectasis
progresses \citep{Sagar2020,Revelo2025}.

To address the need for up-to-date 3D anatomic information during image-guided interventions, we propose \emph{just-in-time volumetric imaging}: a patient-specific learning framework wherein 2D--3D thoracic motion models are learned from pre-intervention scans and completed before they are needed, so that 3D images can be instantaneously inferred from 2D fluoroscopic projections (Figure~\ref{fig:fig1}). This framework generates 3D images in real-time as 2D fluoroscopy is acquired without the need to halt the procedure for additional spins. Importantly, the model is not required during navigation, where robotic-assistance and CBCT-guidance already achieve excellent tool-in-lesion rates; rather, it is deployed at the sampling stage, where repeat spins are typically performed and targeting accuracy marks the difference between tissue that leads to reliable diagnostic decisions and that which does not.

Our contributions are as follows: (i) We formulate volumetric imaging during the intervention as a \emph{dual-scan} problem, in which an inspiration-hold scan supplies a known reference and a ventilated scan supplies the data required to learn patient-specific motion. (ii) Leveraging a known reference scan, we show that the bilinearity of a B-spline motion decomposition renders the 2D--3D problem solvable by alternating least squares (ALS). (iii) We propose the \emph{VM-DS} optimization regime, which trains in minutes and amortizes the model via a projection encoder trained on synthesized 4D data. (iv) Starting from \emph{DREME} \citep{shao2025dreme} as a published baseline, we perform an ablation study to separate the contribution of the reference, the solver and cost function. (v) We evaluate on held-out cases of a public 4DCT dataset and examine model performance as the number of acquired projections varies.

\section{Methods}
\label{sec:methods}

\subsection{Problem statement}
Given reference 3D image $\bm{I}_{\mathrm{ref}} : \Om \subset \R^3 \to \R$ and target 3D image
$\bm{I}(\cdot, p) : \Om \subset \R^3 \to \R$ on shared grid $\Om$, we wish to obtain the deformation vector field (DVF) $\varphi(\x, p)$ such that:
\begin{equation}
  \bm{I}_{\mathrm{ref}}\!\left(\x + \varphi(\x, p)\right) \approx \bm{I}(\x, p)
  \label{eq:refdrr}
\end{equation}
for all $\x \in \Om$. Here $\bm{I}_{\mathrm{ref}}$ is an inspiration-hold CBCT and $\bm{I}(\cdot, p)$ is
the imagined volumetric image capturing the evolving intra-procedural anatomy at the time projection $p$ is acquired, which differs from $\bm{I}_{\mathrm{ref}}$ predominantly by respiratory
motion. Intra-procedural fluoroscopic imaging yields only 2D projections
$p = \proj{\theta}[\bm{I}(\cdot, p)]$, where $\proj{\theta}[\,\cdot\,]$ is the x-ray forward
projector for gantry angle $\theta$. We wish to estimate $\varphi(\x, p)$
from $p$ such that \eqref{eq:refdrr} holds on $\Om$.

\subsection{Baseline method}
Shao \emph{et al.} considered this problem in the single-scan regime, where a single motion-inclusive CBCT projection set $\bm{p}$ is acquired, and proposed the \emph{DREME} joint 3D reconstruction and motion tracking framework \citep{shao2025dreme}. More specifically, 3D reconstruction is achieved via implicit neural representation (INR) \citep{Mildenhall2022,Tewari2022,Molaei2023}, while motion is modeled via a CNN-based encoder and B-spline-based decoder. The INR learns the implicit continuous function $\x \mapsto \bm{I}_{\mathrm{ref}}(\x)$ from a voxel coordinate $\x$ to the corresponding attenuation coefficient $\bm{I}_{\mathrm{ref}}(\x)$, parameterized by a hash encoder and multi-layer perceptron. Meanwhile, the projection-dependent DVF $\varphi(\x, p)$ for voxel $\x$ and projection $p$ is modeled using the low-rank B-spline formalism:
\begin{equation}
  \varphi(\x, p) \;=\; \sum_{i=1}^{3} \bm{w}_i(p) \times \bm{e}_i(\x),
  \label{eq:motion-model}
\end{equation}
where $\bm{w}_i(p)$ are projection-dependent motion basis component (MBC) scores learned via a CNN-based encoder and $\bm{e}_i(\x)$ are voxel-dependent MBCs, shared across all projections. \emph{DREME} is trained with a progressive curriculum of seven stages split across two spatial resolutions ($4$ and $2$ mm$^3$). The method employs two image similarity metrics, four regularizers, and motion and angle augmentation. The full schedule comprises $6{,}800$ epochs.

\subsection{The dual-scan regime}
\label{sec:regime}
\begin{figure}
    \centering
    \includegraphics[width=1\textwidth]{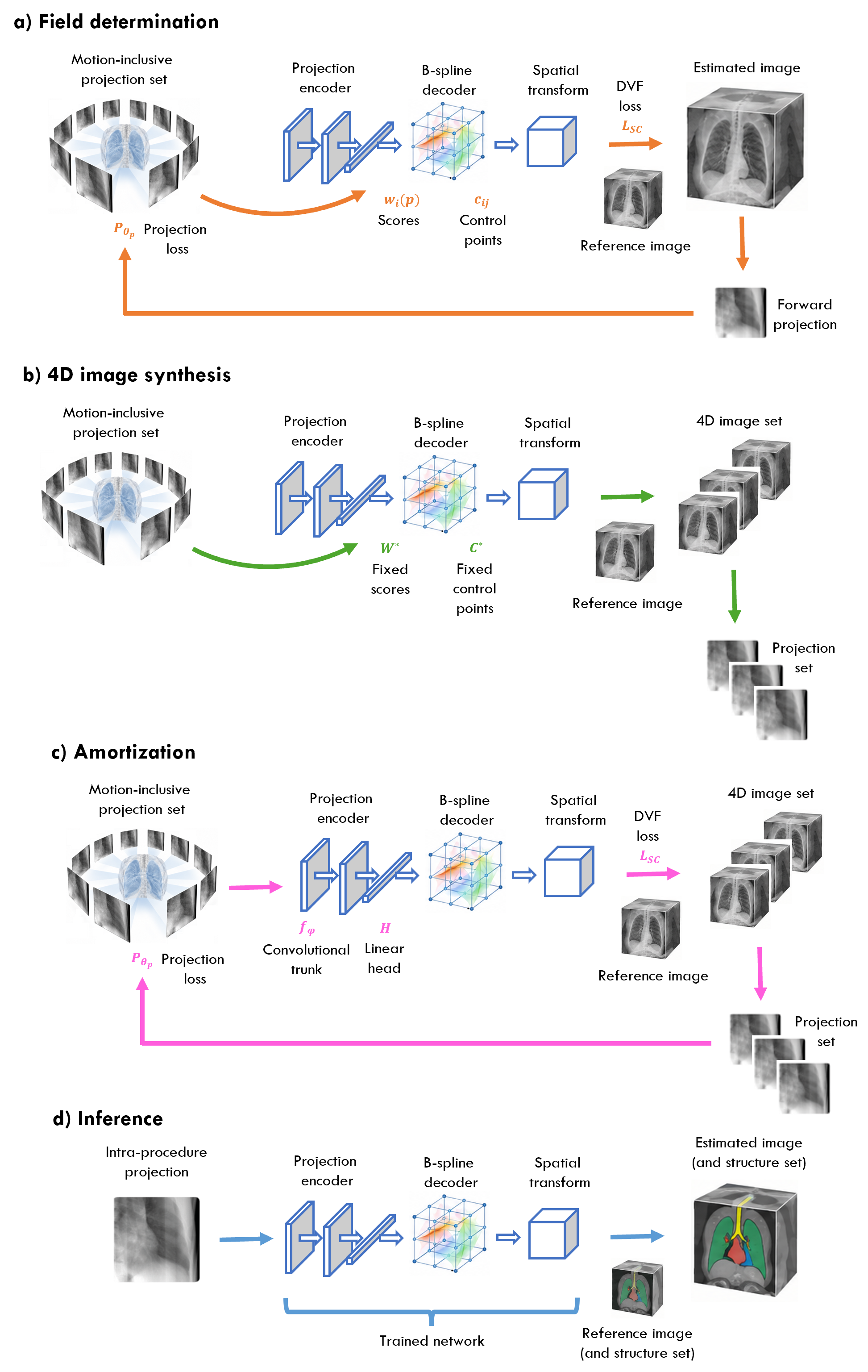}
    \caption{Overview of the \emph{VM-DS} optimization framework. a) During field determination, the scores and control points of the B-spline motion model are solved against the motion-inclusive projections by alternating least squares. b) During synthesis, the solved model warps the inspiration-hold reference to motion states with known scores, which are forward projected at angles across the arc. c) During amortization, the projection encoder is trained on the synthesized pairs alone. d) During inference, the trained network predicts 3D images and structures from intra-procedural projections.}
    \label{fig:fig2}
\end{figure}

While \emph{DREME} learns from a single CBCT, our \emph{VoxelMap} framework originally used a 4D (CT or MRI) dataset \citep{hindley2023,hindley2025} wherein every phase of a 4D scan is forward projected and deformably registered to a reference phase, yielding a training dataset of digitally reconstructed radiographs (DRRs) and DVFs that serve as the inputs and outputs of a CNN. During bronchoscopy a 4D scan is not usually obtained, but an inspiration-hold CBCT $\bm{I}_{\mathrm{ref}}$ is acquired as standard of care to plan the navigation pathway. We therefore acquire one additional motion-inclusive CBCT projection set $\bm{p}$ while the patient is anaesthetised and ventilated, the analogue of \emph{DREME}'s pre-treatment scan (Figure~\ref{fig:fig2}), and estimate motion relative to $\bm{I}_{\mathrm{ref}}$.

A known reference fundamentally alters the nature of the optimization problem. With $\bm{I}_{\mathrm{ref}}$ unknown, the image and the motion trade off against one another such that displacing the image one way and the motion the other leaves every projection unchanged, which \emph{DREME} suppresses with a zero-mean-motion constraint. With $\bm{I}_{\mathrm{ref}}$ known, the only unknowns are the motion model scores and control points. This approach also removes the INR, which accounts for the vast majority of learnable parameters. \emph{DREME} initializes its reference from an FDK reconstruction of the motion-inclusive projections \citep{Feldkamp1984}, which averages the anatomy over the respiratory cycle and blurs it wherever it moves, and motion absorbed into the reference at that stage is not recovered by the later motion estimation (Section~\ref{sec:discussion}). The inspiration-hold CBCT is acquired at a single motion state and carries no such blur.

Ventilation also changes the role of augmentation. Under machine-paced ventilation the respiratory cycle during the procedure repeats the cycle observed by the motion-inclusive scan, so there is no excursion beyond the training range to anticipate and motion-amplitude augmentation can be omitted.

\subsection{\emph{VM-DS}}
\label{sec:vmds}
\subsubsection{Network}
A single shared network is built at the outset,  comprising:
\begin{itemize}
  \item a projection encoder $\mathcal{E}_\psi(\bm{y}_p)$ that has a convolutional trunk $f_\psi$ with weights $\psi$ and a linear head $\bm{H}$, such that $\mathcal{E}_\psi(\bm{y}_p) = \bm{H} f_\psi(\bm{y}_p) = \bm{w}(p) \in \R^{3 \times 3}$ mapping a single projection to motion state scores;
  \item a B-spline decoder that expands the scores into the DVF $\varphi(\x, p)$ through \eqref{eq:bilinear} below; and
  \item a spatial transform module that applies $\varphi$ to $\bm{I}_{\mathrm{ref}}$.
\end{itemize}

\subsubsection{Objective}
Writing the spatial MBCs in \eqref{eq:motion-model} as
$\bm{e}_i(\x) = \sum_{j} \bm{c}_{ij}\,\beta_j(\x)$, where $\beta_j$ are
fixed B-spline basis functions and $\bm{c}_{ij} \in \R^3$ are learnable
control points,
\begin{equation}
  \varphi(\x, p) \;=\; \sum_{i=1}^{3}\sum_{j} \bm{w}_i(p) \odot \bm{c}_{ij}\,\beta_j(\x),
  \label{eq:bilinear}
\end{equation}
which is linear in the scores $\bm{W} = \{\bm{w}_i(p)\}$ when the control
points $\bm{C} = \{\bm{c}_{ij}\}$ are held fixed, and linear in $\bm{C}$
when $\bm{W}$ is held fixed. Since $\bm{I}_{\mathrm{ref}}$ is acquired
rather than learned, $\bm{C}$ and $\bm{W}$ are the only unknowns, and \emph{VM-DS} minimises a single objective:
\begin{equation}
  \mathcal{L}(\varphi) \;=\; \sum_{p \in \bm{p}}
  \Big\{ \tfrac{1}{N_{\mathrm{pixel}}}\!\!\sum_{N_{\mathrm{pixel}}}\!\!
  \rho\!\left(\proj{\theta_p}\!\left[\bm{I}_{\mathrm{ref}}(\x + \varphi(\x,p))\right] - \bm{y}_p\right)
  \;+\; \lambda_{\mathrm{SC}}\, L_{\mathrm{SC}}(p) \Big\},
  \label{eq:vmds-objective}
\end{equation}
where $\bm{I}_{\mathrm{ref}}(\x + \varphi(\x,p))$ is the deformed volume and
$\bm{y}_p$ is the target projection.

The first term is the data term, with $\rho$ a pseudo-Huber penalty on the
projection residual,
\begin{equation}
  \rho(r) \;=\; \frac{r^2}{\sqrt{1 + (r/\delta)^2}},
  \label{eq:rho}
\end{equation}
applied pixelwise. $\rho$ behaves quadratically for $|r| \ll \delta$ and as
$\delta|r|$ for $|r| \gg \delta$, limiting the influence of outlier pixels, and
is written as a reweighted square, $r \mapsto r\,(1 + (r/\delta)^2)^{-1/4}$
with the weight held fixed within a linearisation, so that
\eqref{eq:vmds-objective} remains a sum of squares and the same residual
rows serve both blocks. The transition scale $\delta$ is set to the root
mean square residual at the start of each outer iteration and held fixed
throughout. \emph{DREME} uses the squared error, $\rho(r) = r^2$ \citep{shao2025dreme}, which we retain as an ablation (Section~\ref{sec:ladder}).

The second is a self-consistency term (also employed in \emph{DREME} \citep{shao2025dreme}). Deforming the reference by $\varphi$ and then
by its inverse must return the reference,
\begin{equation}
  L_{\mathrm{SC}}(p) \;=\; \frac{1}{N_{\mathrm{voxel}}} \sum_{l=1}^{N_{\mathrm{voxel}}}
  \Big| \tilde{\bm{I}}_p\!\left(\x_l + \varphi^{-1}(\x_l, p)\right)
        - \bm{I}_{\mathrm{ref}}(\x_l) \Big|^2 ,
  \label{eq:vmds-sc}
\end{equation}
where $\tilde{\bm{I}}_p(\x) = \bm{I}_{\mathrm{ref}}(\x + \varphi(\x,p))$ is the deformed reference and $\varphi^{-1}$ is obtained by three fixed-point iterations
\citep{Chen2008}. This regularizer is evaluated in the volume, penalizing
fields that are not invertible and helping to constrain the ill-posed 2D-3D problem. In the original \emph{VoxelMap} framework, invertibility was encouraged via scaling-and-squaring layers that are slower to compute \citep{hindley2023,hindley2025}.

\subsubsection{Stage 1: Field determination}
In Stage $1$ the projection encoder is frozen, and the
B-spline decoder is solved against the measured projections with the scores
treated as free per-view variables. Rather than descending $\mathcal{L}$ jointly in $(\bm{W}, \bm{C})$, we
alternate between the two blocks, updating one while the other is held
fixed. Each outer iteration consists of a \emph{score step}, \emph{basis step}, and \emph{rescaling} (Algorithm \ref{alg:fdet}).

\begin{algorithm}[H]
\caption{Field determination}
\label{alg:fdet}
\begin{algorithmic}[1]
\Require Reference volume $\bm{I}_{\mathrm{ref}}$; projections $\bm{y}_p$ with gantry angles $\{\theta_p\}$; outer iterations $T$; LM iterations $M$; basis steps $K$; microbatch size $B$; decoding layer $\bm{C}$ and spatial transform
\State Initialise $\bm{C}^{(0)}$: SI-polynomial fields (constant, linear, quadratic), unit Gram norm per axis
\State $\bm{W}^{(0)} \gets \bm{0}$
\For{$t = 1, \dots, T$}
  \State $\delta \gets$ RMS of current projection residuals
  \ForAll{$p \in \bm{p}$ \textbf{in parallel}} \Comment{score step}
    \For{$m = 1, \dots, M$}
      \State Propose $\bm{w}(p)$ by a damped $9 \times 9$ Gauss--Newton solve
      \State Accept if $\|\bm{r}_p\|$ decreases, else increase damping
    \EndFor
  \EndFor
  \State Rescale: $s_{i,k} \gets \|e_{i,k}\|$, \; $e_{i,k} \gets e_{i,k}/s_{i,k}$, \; $w_{i,k} \gets s_{i,k}\, w_{i,k}$
  \For{$k = 1, \dots, K$} \Comment{basis step}
    \State Sample subset $\mathcal{B} \subset \bm{p}$, $|\mathcal{B}| = \min(4B, |\bm{p}|)$
    \State Update $\bm{C}$ by one Adam step on $\nabla_{\bm{C}} \mathcal{L}_{\mathcal{B}}(\bm{W}, \bm{C})$
  \EndFor
  \State Rescale as above
\EndFor
\Ensure Control points $\bm{C}^\star$; per-view scores $\bm{W}^\star$
\end{algorithmic}
\end{algorithm}

\paragraph{Score step}
With $\bm{C}$ fixed, the scores $\bm{w}_i(p)$ of a given projection affect
only its residual. The score step therefore decouples into
independent problems, one per projection, each with nine unknowns
constrained by every detector pixel, i.e.\ of order $10^4$--$10^5$
equations. We solve each by damped Gauss--Newton
(Levenberg--Marquardt) iterations on \eqref{eq:vmds-objective}, with the
Jacobian formed by forward differences and damping scaled by $\mathrm{diag}(\bm{J}^\top\bm{J})$, so
that each iteration requires only a $9 \times 9$ linear solve. Projections are batched and solved in parallel. Both terms
of \eqref{eq:vmds-objective} are carried in the residual, the
self-consistency term evaluated on a fixed random subset of voxels drawn
once, so that the two blocks descend the same function and the residual
remains a deterministic function of $\bm{w}$.

Scores are initialised to zero motion, $\bm{W} = \bm{0}$. As the gradient of
$\mathcal{L}$ with respect to $\bm{C}$ vanishes at $\bm{W} = \bm{0}$, the
control points must be initialised to nonzero fields. Here, we use polynomials in
the normalised SI lattice coordinate $z \in [-1, 1]$ as respiratory
motion predominantly occurs along this axis. For levels $i = 1, 2, 3$, each
Cartesian component of $\bm{e}_i$ is $1$, $z$ and $z^2 -
\overline{z^2}$ respectively, normalised per axis to unit Gram norm, i.e.\
$1$~mm RMS displacement over $\Om$.

\paragraph{Basis step}
With $\bm{W}$ fixed, the control points are updated by $K$ Adam steps on
\eqref{eq:vmds-objective}, each evaluated over a subset of
$\min(4B, |\bm{p}|)$ projections resampled at each outer iteration, with the
gradient accumulated over microbatches of $B$ views to conserve memory. The self-consistency
term is evaluated on the full grid in this step.

\paragraph{Rescaling}
Since the product in \eqref{eq:bilinear} is elementwise, the motion model is
invariant to rescaling any Cartesian component of a given level, $w_{i,k}
\mapsto a\,w_{i,k}$, $e_{i,k} \mapsto e_{i,k}/a$ and the optimizer can drift along these directions without changing $\mathcal{L}$. Therefore, after each solve, every component $e_{i,k}$ is rescaled to unit norm and the corresponding score component $w_{i,k}$
is rescaled inversely, leaving $\varphi$ unchanged.

\subsubsection{Stage 2: Synthesis and amortization}
At the end of Stage $1$, the control points and scores are frozen at $\bm{C}^\star$ and $\bm{W}^\star$ and the decoder is used to generate training data with \emph{known} motion. We draw $N_{\mathrm{s}}$ score vectors directly from the measured projections, each by interpolating at a random fraction between the solved scores of two consecutive training projections in acquisition order, warp $\bm{I}_{\mathrm{ref}}$ by the corresponding fields \eqref{eq:bilinear}, and forward project each warped volume at a gantry angle drawn uniformly from the acquisition arc (Algorithm~\ref{alg:amor}). Each synthesized projection $\hat{\bm{y}}_s$ is thus paired with the exact scores $\bm{w}_s$ that produced it.

The encoder, trunk and linear head together, is trained by Adam on the score regression $\sum_s \|\mathcal{E}_\psi(\hat{\bm{y}}_s) - \bm{w}_s\|^2$ over the synthesized pairs, and never takes a gradient step on measured data. This matters for fluoroscopy at a fixed C-arm angle. On a single breath every measured angle carries one respiratory phase, so an encoder fitted to the measured projections can learn the phase from the angle, which a fixed-angle sequence does not supply. Therefore, the synthesized pairs provide every motion state with every angle. With $N_{\mathrm{s}} = 4096$ pairs and $E_{\mathrm{s}} = 100$ epochs the encoder sees about $4 \times 10^5$ projections, half the exposure of the \emph{DREME} encoder ($3{,}700$ epochs of stages I-c, II-c and II-d over $227$ projections), and Stage $2$ takes approximately $43$~s. $E_{\mathrm{s}}$ was selected on development case~$8$ (Section~\ref{sec:config-methods}).

\begin{algorithm}[H]
\caption{Synthesis and amortization}
\label{alg:amor}
\begin{algorithmic}[1]
\Require Reference volume $\bm{I}_{\mathrm{ref}}$; measured projections $\bm{y}_p$; Stage $1$ output $\bm{C}^\star$, $\bm{W}^\star$; number of pairs $N_{\mathrm{s}}$; epochs $E_{\mathrm{s}}$; trunk $f_\psi$, head $\bm{H}$
\For{$s = 1, \dots, N_{\mathrm{s}}$}
  \State Draw consecutive training projections $(p, p')$ in acquisition order, $\tau \sim U[0,1]$, angle $\theta_s \sim U(\mathrm{arc})$
  \State $\bm{w}_s \gets (1-\tau)\,\bm{w}^\star(p) + \tau\,\bm{w}^\star(p')$
  \State $\hat{\bm{y}}_s \gets \proj{\theta_s}[\bm{I}_{\mathrm{ref}}(\cdot + \varphi(\cdot\,; \bm{w}_s, \bm{C}^\star))]$
\EndFor
\State Initialise $\psi$, $\bm{H}$; for $E_{\mathrm{s}}$ epochs update both by Adam on $\sum_s \|\bm{H} f_\psi(\hat{\bm{y}}_s) - \bm{w}_s\|^2$
\Ensure Network: trunk $f_\psi$, head $\bm{H}$, decoding layer $\bm{C}^\star$, spatial transform
\end{algorithmic}
\end{algorithm}

\subsubsection{Inference}
The network that runs at inference comprises the convolutional trunk $f_\psi$, linear head $\bm{H}$, decoding layer $\bm{C}^\star$ and spatial transform. Given a single intra-procedural fluoroscopic projection $\bm{y}$, one forward pass predicts the scores $\bm{w} = \mathcal{E}_\psi(\bm{y})$, expands them through \eqref{eq:bilinear} into the DVF $\varphi(\x)$, and warps $\bm{I}_{\mathrm{ref}}$ to produce the volumetric image $\bm{I}_{\mathrm{ref}}(\x + \varphi(\x))$. No optimization runs during the intervention. At batch size one this forward pass takes $1.49$~ms (median of $500$ timed passes; $95$th percentile $1.53$~ms), of which the encoder accounts for $0.13$~ms, the decoding layer $0.66$~ms and the warp $0.68$~ms.

\subsection{Comparators and component ladder}
\label{sec:ladder}
\emph{DREME} reconstructs its reference from the motion-inclusive scan alone and trains its seven-stage curriculum ($6{,}800$ epochs) without a time limit. In the dual-scan regime the inspiration-hold image is available to any method, so the comparator for the held-out cases, referred to simply as \emph{DREME}, is that curriculum given the inspiration-hold image as its reference, fitted by the INR and then held fixed. \emph{DREME}$_{fdk}$ and \emph{DREME}$_{Iref}$ are included here only for completeness. Given the dual-scan regime with a known reference image, learning only on the motion-inclusive CBCT or allowing the reference to absorb motion would unfairly disadvantage the method, especially because the landmarks used for evaluation are defined per phase, hence averaging across phases leads to smearing. To highlight the differences between \emph{DREME} as a published baseline and \emph{VM-DS}, we vary one component at a time (Table~\ref{tab:ladder}):
\begin{itemize}
  \item \textbf{\emph{DREME}$_{fdk}$}: as published, with the INR reference learned from an FDK reconstruction of the motion-inclusive scan \citep{Feldkamp1984}.
  \item \textbf{\emph{DREME}$_{Iref}$}: the same curriculum with the INR initialized from the inspiration-hold reference and then learned with the motion.
  \item \textbf{\emph{DREME}}: the INR fitted to the inspiration-hold reference and then held fixed, so that the reference cannot absorb motion. This isolates the dual-scan regime and is the held-out comparator.
  \item \textbf{\emph{DREME}$_{-aug}$}: motion and angle augmentation removed from every stage, so the final, augmentation-only stage drops out (Section~\ref{sec:regime}).
  \item \textbf{\emph{DREME}$_{-aug,-INR}$}: the INR removed and the inspiration-hold reference used directly, so the stages that only fit or refine the reference drop out. Two stages remain, both pure motion estimation by stochastic gradient descent (SGD).
  \item \textbf{\emph{VM-DS}$_{SSD}$}: the SGD curriculum replaced with field determination by ALS, 4D image synthesis, and amortization, within a $10$-minute window, with a squared-error data term.
  \item \textbf{\emph{VM-DS}}: the pseudo-Huber data term \eqref{eq:rho} in place of the squared error. A subscript gives the stride $s$ of the solved projections, so that \emph{VM-DS}$_{s=3}$ solves on every third projection and \emph{VM-DS}$_{s=12}$ on every twelfth.
  \item \textbf{\emph{VM-DS}$_{-SC}$}: self-consistency term removed ($\lambda_{\mathrm{SC}} = 0$).
\end{itemize}

\subsection{Data}
We evaluate on the public DIR-Lab thoracic 4DCT dataset \citep{Castillo2009}, which provides ten cases at ten respiratory phases (T00--T90), with $300$ expert landmarks at the extreme phases and $75$ at six phases. Each phase volume is resampled to $1 \times 1 \times 1$~mm$^3$, cropped to a single box of $288$~(LR) $\times\ 320$~(SI) $\times\ 248$~(AP)~mm sized to contain every annotated DIR-Lab landmark, and downsampled to $2 \times 2 \times 2$~mm$^3$. The documented end-inhale phase represents the inspiration-hold image and supplies $\bm{I}_{\mathrm{ref}}$.

The motion-inclusive projection set $\bm{p}$ is simulated as a circular cone-beam arc of $680$ projections over $360^{\circ}$, with source-to-axis and source-to-detector distances of $1000$ and $1500$~mm and a $272 \times 320$ detector at $1.8$~mm pitch, so that one detector pixel subtends $1.2$~mm at isocentre. As anaesthetised ventilation is machine-paced, respiration is modelled as a triangular cycle across the arc, with each projection rendered from the phase volume of its respiratory bin by ray-driven forward projection.

DIR-Lab cases~1 and~8, the lowest- and highest-motion cases, were used for development and are reported separately. All hyperparameters were determined on these two cases (Section~\ref{sec:config-methods}), and the held-out cases were each run once at the resulting fixed configuration.

\subsection{Hyperparameter search}
\label{sec:config-methods}
The $10$-minute window is treated as a hard budget, so Stage~1 hyperparameters were selected for accuracy at this fixed cost rather than individually. Stage~1 is capped at $8.5$~minutes so that the complete procedure, including setup and Stage~2, fits within $10$~minutes. Its cost is set by the Gauss--Newton iterations $M$, the outer iterations $T$, the basis steps per outer iteration $K$ and the number of solved views. Measuring  per-step costs provides a cost model under which, for fixed $M$ and solved-view stride, $T$ and $K$ exchange along an iso-cost line that fills the cap. Fewer solved views make each outer iteration cheaper and move the line outward. Cells along each line, at solved-view strides $s \in \{3, 6, 12, 30\}$ ($227$, $114$, $57$ and $23$ views), were run on the development cases, each followed by the same Stage~2, and scored by held-out TRE on the arc and at a fixed anterior--posterior angle (Section~\ref{sec:eval}). The number of solved projections is an important hyperparameter. As the arc covers a single breath, a sparser arc also samples that breath more coarsely, but buys more outer iterations within the window. The configuration was selected on case~$8$, whose large motion separates the cells, with case~$1$ as a check, and fixed for all held-out cases. The Stage~2 epochs $E_{\mathrm{s}}$ were chosen the same way, on case~$8$ with $N_{\mathrm{s}} = 4096$. A fixed budget is used throughout rather than adaptive stopping, as the training residual flattens well before TRE does and a TRE-based stopping rule would use ground truth not available in normal clinical settings.

\subsection{Evaluation}
\label{sec:eval}
Consistent with \emph{DREME}, every third projection is available for training, here $227$ views by default, and the remaining $453$ views are held out for testing at every configuration. Accuracy is reported as landmark target registration error (TRE). Let $\{\x_n\}_{n=1}^{N_L}$ be the landmarks annotated on the reference phase and $\{\x_n^{b}\}$ their annotated positions at phase bin $b$. A feature at $\x_n$ in the reference appears at $\x_n + \varphi^{-1}(\x_n, p)$, and the error at one landmark per view is
\begin{equation}
  TRE(p) \;=\; \big\| \x_n + \varphi^{-1}(\x_n, p) - \x_n^{b(p)} \big\|_2 ,
  \label{eq:tre-point}
\end{equation}
where $b(p)$ is the phase bin of projection $p$ and $\varphi$ is evaluated at the encoder-predicted scores. TRE is averaged over the landmarks and over every held-out view in a phase, and reported as the mean over the moving phases with landmarks. Image similarity between the predicted volume and the ground-truth phase volume is assessed by the structural similarity index (SSIM) and the peak signal-to-noise ratio (PSNR), evaluated within a body mask on the $2$~mm grid over every held-out view.

As the single-breath arc couples gantry angle and respiratory phase, TRE is reported in two ways. On the \emph{arc}, each held-out view is predicted at the phase at which it was acquired. At \emph{fixed AP}, as in fluoroscopy at a fixed C-arm angle, the held-out views within $\pm 10^\circ$ of the anterior--posterior direction are rendered at every moving phase, so that each phase is seen at an angle where the training arc did not acquire it. Robustness to quantum noise is assessed for \emph{VM-DS}$_{s=12}$ alone by repeating the held-out evaluation with Poisson noise applied to every projection, counts $N \sim \mathrm{Poisson}(I_0\, e^{-\ell})$ for line integral $\ell$, returned as $-\ln(N/I_0)$, at $I_0 = 10^5$ and $10^4$ unattenuated photons per detector pixel.

\subsection{Implementation details}
To isolate the effect of the optimization regime, the projection encoder, the B-spline decoder and the self-consistency weight are all held constant. The encoder takes a single $272 \times 320$ projection and applies six layers of $3 \times 3$ convolution, each followed by batch normalization and ReLU, with $2$, $4$, $8$, $16$, $32$ and $32$ channels, followed by a linear layer from the $32 \times 5 \times 5$ features to the nine scores ($21{,}405$ parameters, of which the linear head has $5{,}769$). The decoder has three spatial levels of uniform cubic B-splines, one motion basis component per level, the lattice doubling at each level with $8$, $14$ and $26$ control points per axis at the three levels. The self-consistency term is weighted by $\lambda_{\mathrm{SC}} = 10^3$, as in \emph{DREME}. The unit-norm constraint on the basis components \citep[Eq.~6]{shao2025dreme} is evaluated as a mean over voxels in millimetres, so that one unit of score corresponds to $1$~mm root-mean-square displacement. In Stage~1, field determination uses batches of $B = 32$ projections and an Adam learning rate of $3 \times 10^{-3}$ for the control points. In Stage~2 the encoder is trained for $100$ epochs on $4096$ synthesized pairs by Adam at a learning rate of $10^{-3}$. All experiments ran on a workstation with a NVIDIA RTX PRO 4000 Blackwell GPU and an AMD Ryzen Threadripper PRO 9965WX CPU, using PyTorch~2.7~\cite{Paszke2019} and the ASTRA toolbox~2.5~\cite{vanAarle2015,vanAarle2016} for forward projection.

\section{Results}

\subsection{Configuration search}
\label{sec:config}
\begin{figure}[p]
    \centering
    \includegraphics[width=1\textwidth]{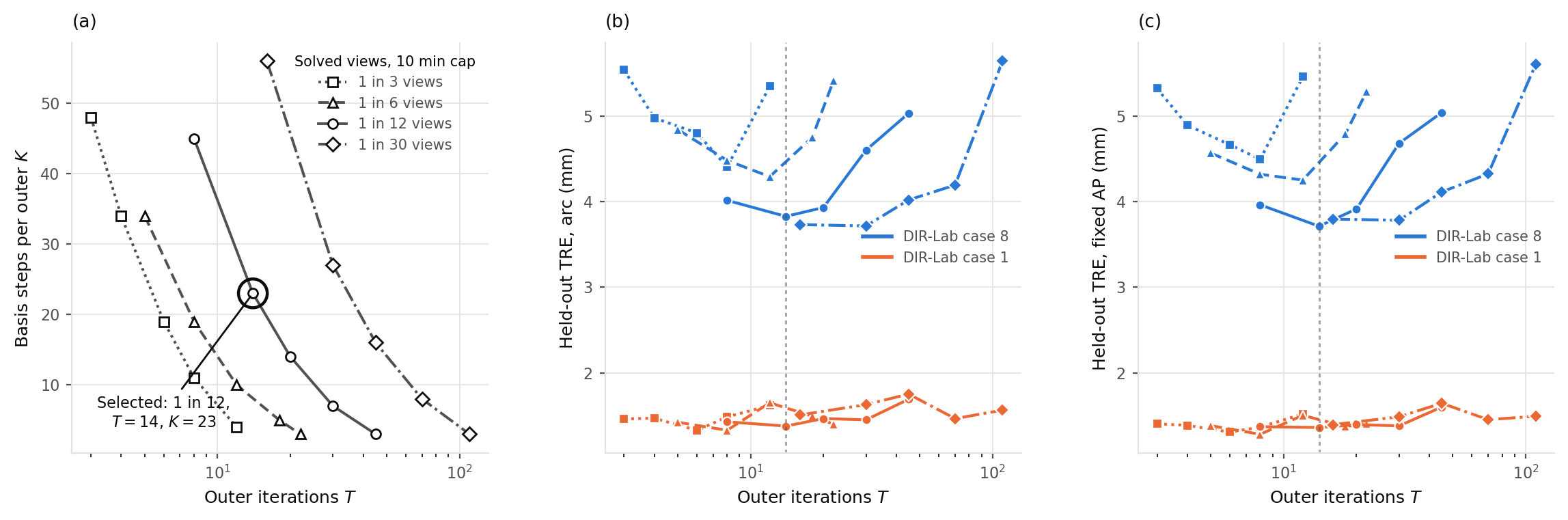}
    \caption{Configuration search on the development cases within the $10$-minute window. (a) Iso-cost lines in the plane of outer iterations $T$ and basis steps per outer iteration $K$. Held-out TRE along each line on the arc (b) and at fixed AP (c), on case~$8$ (blue, high motion) and case~$1$ (orange, low motion).}
    \label{fig:commission}
\end{figure}

Figure~\ref{fig:commission} reports the Stage~1 configuration search on the development cases (Section~\ref{sec:config-methods}), every cell followed by the same Stage~2. The selected configuration is $M = 1$, solving on every twelfth projection ($57$ views) with $T = 14$ outer iterations of $K = 23$ basis steps. It gave $3.8$~mm on the arc and $3.7$~mm at fixed AP on case~$8$, and $1.4$~mm on both on case~$1$, the lowest at fixed AP on case~$8$ and within $0.1$~mm of the lowest elsewhere. At its stride, cells with fewer basis steps per outer iteration were less accurate on case~$8$ ($4.6$~mm at $T = 30$, $K = 7$) and on both cases at $T = 45$, $K = 3$. Arc and fixed-AP TRE agreed to within $0.15$~mm at every cell. Stage~1 used $8.5$ and $8.7$~minutes on cases~$1$ and~$8$, the cap being checked between outer iterations, and Stage~2 took $43$~s.

\begin{figure}[htbp]
    \centering
    \includegraphics[width=1\textwidth]{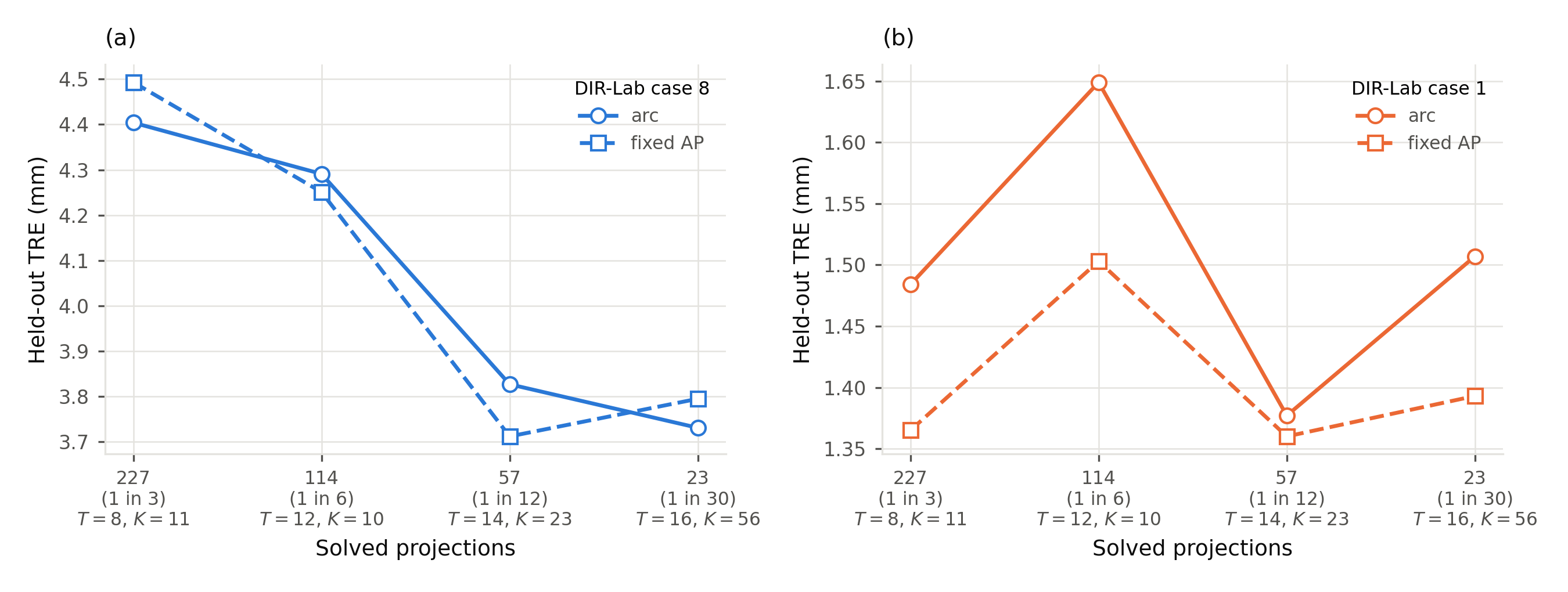}
    \caption{Projection number on the development cases. Held-out TRE on the arc and at fixed AP for case~$8$ (a) and case~$1$ (b) against the number of solved projections, each at the cell of its iso-cost line with the lowest case-$8$ TRE over all held-out angles and phases (Figure~\ref{fig:commission}), followed by the same Stage~2.}
    \label{fig:projnum}
\end{figure}

Figure~\ref{fig:projnum} shows the effect of projection number at fixed cost. With $227$ and $114$ solved views the window admits too few outer iterations, and case~$8$ TRE was $4.4$ and $4.3$~mm on the arc; with $57$ and $23$ views it was $3.8$ and $3.7$~mm, against $3.8$~mm for \emph{DREME} solving on $227$. On case~$1$ every count gave $1.4$--$1.7$~mm. Within the window, fewer solved views were therefore no less accurate, and the held-out evaluation reports both $57$ (\emph{VM-DS}$_{s=12}$) and $23$ (\emph{VM-DS}$_{s=30}$, $T = 16$, $K = 56$, the lowest stride-$30$ cell on case~$8$).

Stage~2 exposure was selected on case~$8$, before the configuration search, on its Stage~1 model at stride $12$, $T = 20$, $K = 14$. With $256$ synthesized pairs, TRE over all held-out angles and phases was $4.3$~mm whether the encoder was trained for $200$ or $800$ epochs, whereas $4096$ pairs gave $4.1$, $3.9$, $3.8$ and $3.8$~mm at $25$, $50$, $100$ and $200$ epochs, against $3.8$~mm for \emph{DREME}; $16{,}384$ pairs gave no further gain. The number of distinct synthesized motion states, rather than the number of passes over them, limited the encoder, and $N_{\mathrm{s}} = 4096$, $E_{\mathrm{s}} = 100$ was fixed.

\subsection{Component ladder}
\label{sec:res-ladder}
Table~\ref{tab:ladder} reports the ladder on the development cases. Supplying the inspiration-hold reference as the initial INR (\emph{DREME}$_{Iref}$) reduced TRE from $11.0$ to $7.8$~mm on case~$8$ and from $2.5$ to $2.1$~mm on case~$1$, at a training time of $398$~minutes. Holding the INR fixed (\emph{DREME}) reduced TRE from $11.0$ to $3.8$~mm on case~$8$ and to $1.4$~mm on case~$1$ in $173$~minutes. Removing augmentation and then the INR (\emph{DREME}$_{-aug}$, \emph{DREME}$_{-aug,-INR}$) both had negligible impact on TRE while shortening training to $135$~minutes. Replacing the SGD curriculum by the \emph{VM-DS} workflow in the $10$-minute window, with the encoder trained on synthesized pairs only, gave $4.7$ and $1.7$~mm TRE for cases $8$ and $1$ with the squared error (\emph{VM-DS}$_{SSD}$) and $4.8$ and $1.5$~mm with the pseudo-Huber term (\emph{VM-DS}$_{s=3}$). At the selected configuration, solving on every twelfth projection so that the window admits more outer iterations, \emph{VM-DS}$_{s=12}$ reached $3.7$ and $1.5$~mm, against $3.8$ and $1.4$~mm for \emph{DREME}, and against $4.4$ and $1.9$~mm with the squared error at the same configuration. At fixed AP it gave $3.7$ and $1.4$~mm. Removing $L_{\mathrm{SC}}$ (\emph{VM-DS}$_{-SC}$) raised TRE to $12.5$~mm on case~$8$ and $93.9$~mm on case~$1$, indicating divergence of the optimizer.

\begin{figure}
    \centering
    \includegraphics[width=1\textwidth]{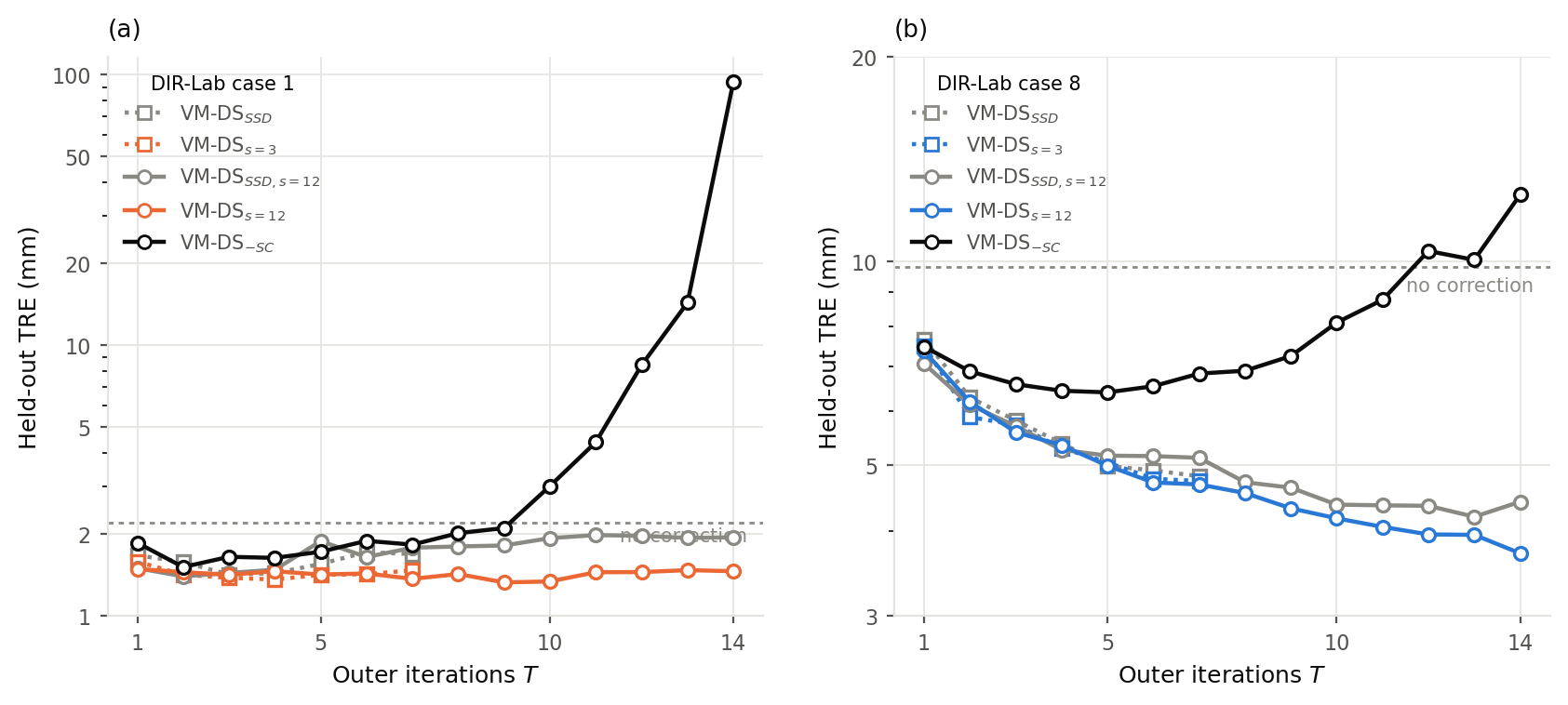}
    \caption{Held-out TRE per outer iteration of the \emph{VM-DS} ladder rows on development case~$1$ (a) and case~$8$ (b). Each point is the Stage~1 model at that iteration followed by the same Stage~2. The dotted line is no correction.}
    \label{fig:ladder-curves}
\end{figure}

Figure~\ref{fig:ladder-curves} shows held-out TRE per outer iteration, each point being the Stage~1 model at that iteration passed through the same Stage~2. With $L_{\mathrm{SC}}$ and the pseudo-Huber term, TRE on case~$8$ fell throughout the window, to $3.7$~mm at $T = 14$, and on case~$1$ it reached $1.4$~mm by the third outer iteration and stayed within $0.1$~mm of it. With the squared error at stride $12$, TRE on case~$1$ rose after the second outer iteration, to $1.9$~mm, and on case~$8$ it fell more slowly, to $4.4$~mm. Without $L_{\mathrm{SC}}$, TRE on case~$8$ reached $6.4$~mm at the fifth outer iteration and then rose, exceeding no correction from the twelfth. On case~$1$ it rose from $1.5$~mm at the second to $93.9$~mm at the fourteenth, as the root-mean-square motion score, where one unit is $1$~mm root-mean-square displacement, grew from $0.2$ to $46$. The Stage~1 training residual stayed at or below that of \emph{VM-DS}$_{s=12}$ for the first nine outer iterations on case~$1$ and the first twelve on case~$8$, by which point TRE had already risen. The data term alone appears to be satisfied by fields that reproduce the projections without reproducing the anatomy, and the residual does not reveal it until the field has diverged.

\begin{table}[htbp]
\centering
\caption{Component ladder on the development cases (held-out views, TRE in mm, mean $\pm$ standard deviation over the moving phases). Each row changes one component of the row above. All rows solve on every third projection except the last two, which use the selected configuration. \emph{VM-DS}$_{-SC}$ runs the same $14$ outer iterations as \emph{VM-DS}$_{s=12}$. \emph{DREME}$_{fdk}$ and \emph{DREME}$_{Iref}$ learn their reference and are scored in their own reference frame.}
\label{tab:ladder}
\small
\begin{tabular}{llccc}
\toprule
method & change & case 8 & case 1 & time (min) \\
\midrule
no correction &  & $9.8 \pm 5.3$ & $2.2 \pm 1.4$ & \\
\emph{DREME}$_{fdk}$ & as published & $11.0 \pm 3.1$ & $2.5 \pm 0.7$ & $400$ \\
\emph{DREME}$_{Iref}$ & + inspiration-hold reference & $7.8 \pm 3.2$ & $2.1 \pm 1.0$ & $398$ \\
\emph{DREME} & INR fixed & $3.8 \pm 1.5$ & $1.4 \pm 0.6$ & $173$ \\
\emph{DREME}$_{-aug}$ & $-$ augmentation & $3.9 \pm 1.5$ & $1.5 \pm 0.6$ & $149$ \\
\emph{DREME}$_{-aug,-INR}$ & $-$ INR & $3.9 \pm 1.4$ & $1.4 \pm 0.6$ & $135$ \\
\emph{VM-DS}$_{SSD}$ & SGD $\to$ ALS, synthesis, amortization & $4.7 \pm 2.0$ & $1.7 \pm 0.8$ & $10$ \\
\emph{VM-DS}$_{s=3}$ & + pseudo-Huber & $4.8 \pm 1.9$ & $1.5 \pm 0.6$ & $10$ \\
\emph{VM-DS}$_{s=12}$ & every twelfth projection & $3.7 \pm 1.4$ & $1.5 \pm 0.6$ & $10$ \\
\emph{VM-DS}$_{-SC}$ & $-$ $L_{\mathrm{SC}}$ & $12.5 \pm 5.8$ & $93.9 \pm 43.7$ & $3$ \\
\bottomrule
\end{tabular}
\end{table}

\subsection{Held-out evaluation}
\label{sec:res-heldout}
Table~\ref{tab:heldout} reports the held-out cases, both on the arc and at a fixed anterior--posterior (AP) angle. With no correction the mean TRE is $5.7$~mm. On the arc, \emph{DREME} reduced it to $2.7$~mm in $173$ minutes of training, and \emph{VM-DS}$_{s=12}$ and \emph{VM-DS}$_{s=30}$ reached $2.8$~mm each within the $10$-minute window. At fixed AP, \emph{DREME} gave $2.6$~mm and \emph{VM-DS}$_{s=12}$ and \emph{VM-DS}$_{s=30}$ gave $2.8$ and $2.7$~mm, within $0.1$~mm of their arc TRE. No \emph{VM-DS} configuration differed significantly from \emph{DREME} on the arc ($+0.1$~mm, $p = 0.64$ and $0.55$) or at fixed AP ($+0.2$ and $+0.1$~mm, $p = 0.25$ and $0.38$); \emph{DREME} was more accurate on the arc on five of the eight cases against either. Solving on $23$ rather than $57$ projections changed the mean by less than $0.05$~mm on the arc and at fixed AP ($p = 0.31$ and $0.46$).

\begin{table}[htbp]
\centering
\caption{Held-out TRE (mm) on the DIR-Lab cases. Per case, mean $\pm$ standard deviation over the moving phases; the \emph{vs} rows give the paired mean difference with the Wilcoxon signed-rank $p$ in parentheses.}
\label{tab:heldout}
\small
\setlength{\tabcolsep}{3pt}
\resizebox{\textwidth}{!}{%
\begin{tabular}{lccccccc}
\toprule
 & & \multicolumn{2}{c}{\emph{DREME}} & \multicolumn{2}{c}{\emph{VM-DS}$_{s=12}$} & \multicolumn{2}{c}{\emph{VM-DS}$_{s=30}$} \\
\cmidrule(lr){3-4}\cmidrule(lr){5-6}\cmidrule(lr){7-8}
case & no correction & arc & fixed AP & arc & fixed AP & arc & fixed AP \\
\midrule
2 & $3.8 \pm 1.1$ & $1.8 \pm 0.2$ & $1.8 \pm 0.1$ & $1.9 \pm 0.4$ & $1.8 \pm 0.3$ & $1.9 \pm 0.4$ & $1.9 \pm 0.3$ \\
3 & $5.1 \pm 2.1$ & $1.9 \pm 0.3$ & $1.9 \pm 0.3$ & $2.5 \pm 0.6$ & $2.5 \pm 0.5$ & $2.5 \pm 0.6$ & $2.3 \pm 0.5$ \\
4 & $6.7 \pm 2.6$ & $3.1 \pm 0.8$ & $2.8 \pm 0.6$ & $2.9 \pm 0.8$ & $2.7 \pm 0.6$ & $2.6 \pm 0.5$ & $2.6 \pm 0.5$ \\
5 & $5.2 \pm 2.0$ & $2.9 \pm 0.3$ & $2.9 \pm 0.3$ & $3.0 \pm 0.8$ & $3.1 \pm 0.8$ & $3.0 \pm 0.8$ & $3.1 \pm 0.7$ \\
6 & $7.4 \pm 3.5$ & $3.8 \pm 0.6$ & $3.6 \pm 0.8$ & $3.5 \pm 0.7$ & $3.7 \pm 1.0$ & $3.7 \pm 1.0$ & $3.6 \pm 0.8$ \\
7 & $6.7 \pm 3.1$ & $2.5 \pm 0.6$ & $2.6 \pm 0.6$ & $3.2 \pm 0.9$ & $3.1 \pm 0.9$ & $3.3 \pm 0.8$ & $3.2 \pm 0.7$ \\
9 & $5.0 \pm 2.2$ & $2.9 \pm 0.3$ & $2.6 \pm 0.3$ & $2.4 \pm 0.6$ & $2.3 \pm 0.6$ & $2.4 \pm 0.6$ & $2.4 \pm 0.6$ \\
10 & $5.4 \pm 2.6$ & $2.7 \pm 0.7$ & $2.6 \pm 0.7$ & $2.9 \pm 0.8$ & $2.9 \pm 0.7$ & $2.9 \pm 0.7$ & $2.8 \pm 0.7$ \\
\midrule
mean & $5.7 \pm 1.1$ & $2.7 \pm 0.6$ & $2.6 \pm 0.6$ & $2.8 \pm 0.5$ & $2.8 \pm 0.6$ & $2.8 \pm 0.6$ & $2.7 \pm 0.5$ \\
\midrule
vs \emph{DREME} & & & & $+0.1$ ($0.64$) & $+0.2$ ($0.25$) & $+0.1$ ($0.55$) & $+0.1$ ($0.38$) \\
vs \emph{VM-DS}$_{s=12}$ & & & & & & $+0.0$ ($0.31$) & $-0.0$ ($0.46$) \\
\midrule
training time (min) & & \multicolumn{2}{c}{$173$} & \multicolumn{2}{c}{$10$} & \multicolumn{2}{c}{$10$} \\
\bottomrule
\end{tabular}}
\end{table}

\begin{table}[htbp]
\centering
\caption{Held-out SSIM on the DIR-Lab cases. Per case, mean $\pm$ standard deviation over the moving phases; the \emph{vs} rows give the paired mean difference with the Wilcoxon signed-rank $p$ in parentheses.}
\label{tab:ssim}
\small
\setlength{\tabcolsep}{3pt}
\resizebox{\textwidth}{!}{%
\begin{tabular}{lccccccc}
\toprule
 & & \multicolumn{2}{c}{\emph{DREME}} & \multicolumn{2}{c}{\emph{VM-DS}$_{s=12}$} & \multicolumn{2}{c}{\emph{VM-DS}$_{s=30}$} \\
\cmidrule(lr){3-4}\cmidrule(lr){5-6}\cmidrule(lr){7-8}
case & no correction & arc & fixed AP & arc & fixed AP & arc & fixed AP \\
\midrule
2 & $0.89 \pm 0.04$ & $0.95 \pm 0.01$ & $0.95 \pm 0.00$ & $0.95 \pm 0.01$ & $0.95 \pm 0.01$ & $0.95 \pm 0.01$ & $0.95 \pm 0.01$ \\
3 & $0.97 \pm 0.02$ & $0.99 \pm 0.00$ & $0.99 \pm 0.00$ & $0.99 \pm 0.00$ & $0.99 \pm 0.00$ & $0.99 \pm 0.00$ & $0.99 \pm 0.00$ \\
4 & $0.92 \pm 0.03$ & $0.96 \pm 0.01$ & $0.96 \pm 0.01$ & $0.96 \pm 0.01$ & $0.96 \pm 0.01$ & $0.96 \pm 0.01$ & $0.95 \pm 0.01$ \\
5 & $0.93 \pm 0.03$ & $0.97 \pm 0.00$ & $0.97 \pm 0.00$ & $0.96 \pm 0.01$ & $0.96 \pm 0.01$ & $0.96 \pm 0.01$ & $0.96 \pm 0.01$ \\
6 & $0.90 \pm 0.04$ & $0.93 \pm 0.02$ & $0.93 \pm 0.02$ & $0.93 \pm 0.02$ & $0.94 \pm 0.02$ & $0.93 \pm 0.02$ & $0.93 \pm 0.02$ \\
7 & $0.89 \pm 0.04$ & $0.95 \pm 0.01$ & $0.95 \pm 0.01$ & $0.93 \pm 0.02$ & $0.94 \pm 0.01$ & $0.93 \pm 0.02$ & $0.93 \pm 0.01$ \\
9 & $0.92 \pm 0.03$ & $0.96 \pm 0.01$ & $0.96 \pm 0.01$ & $0.96 \pm 0.01$ & $0.96 \pm 0.01$ & $0.96 \pm 0.01$ & $0.96 \pm 0.01$ \\
10 & $0.91 \pm 0.04$ & $0.95 \pm 0.01$ & $0.95 \pm 0.01$ & $0.95 \pm 0.01$ & $0.95 \pm 0.01$ & $0.94 \pm 0.02$ & $0.95 \pm 0.01$ \\
\midrule
mean & $0.92 \pm 0.02$ & $0.96 \pm 0.02$ & $0.96 \pm 0.02$ & $0.95 \pm 0.02$ & $0.96 \pm 0.02$ & $0.95 \pm 0.02$ & $0.95 \pm 0.02$ \\
\midrule
vs \emph{DREME} & & & & $-0.00$ ($0.20$) & $-0.00$ ($0.55$) & $-0.01$ ($0.02$) & $-0.01$ ($0.04$) \\
vs \emph{VM-DS}$_{s=12}$ & & & & & & $-0.00$ ($0.02$) & $-0.00$ ($0.02$) \\
\bottomrule
\end{tabular}}
\end{table}

\begin{table}[htbp]
\centering
\caption{Held-out PSNR (dB) on the DIR-Lab cases. Per case, mean $\pm$ standard deviation over the moving phases; the \emph{vs} rows give the paired mean difference with the Wilcoxon signed-rank $p$ in parentheses.}
\label{tab:psnr}
\small
\setlength{\tabcolsep}{3pt}
\resizebox{\textwidth}{!}{%
\begin{tabular}{lccccccc}
\toprule
 & & \multicolumn{2}{c}{\emph{DREME}} & \multicolumn{2}{c}{\emph{VM-DS}$_{s=12}$} & \multicolumn{2}{c}{\emph{VM-DS}$_{s=30}$} \\
\cmidrule(lr){3-4}\cmidrule(lr){5-6}\cmidrule(lr){7-8}
case & no correction & arc & fixed AP & arc & fixed AP & arc & fixed AP \\
\midrule
2 & $28.8 \pm 3.6$ & $34.4 \pm 0.7$ & $34.6 \pm 0.6$ & $34.4 \pm 1.4$ & $34.5 \pm 1.2$ & $34.3 \pm 1.5$ & $34.4 \pm 1.2$ \\
3 & $42.5 \pm 4.1$ & $49.2 \pm 1.1$ & $49.4 \pm 1.2$ & $47.7 \pm 1.9$ & $47.5 \pm 1.8$ & $47.3 \pm 1.9$ & $47.8 \pm 1.6$ \\
4 & $33.6 \pm 2.8$ & $38.4 \pm 1.8$ & $38.9 \pm 1.5$ & $38.1 \pm 2.1$ & $38.4 \pm 1.8$ & $37.8 \pm 1.8$ & $37.7 \pm 2.1$ \\
5 & $33.0 \pm 2.9$ & $38.9 \pm 0.7$ & $38.6 \pm 0.5$ & $37.4 \pm 1.2$ & $37.4 \pm 1.2$ & $37.6 \pm 1.2$ & $37.5 \pm 1.1$ \\
6 & $32.7 \pm 3.5$ & $35.3 \pm 1.9$ & $35.2 \pm 2.2$ & $35.6 \pm 2.3$ & $35.9 \pm 2.3$ & $34.9 \pm 2.0$ & $35.3 \pm 1.8$ \\
7 & $31.6 \pm 3.0$ & $36.4 \pm 1.4$ & $36.4 \pm 1.2$ & $35.0 \pm 1.8$ & $35.2 \pm 1.3$ & $34.5 \pm 1.6$ & $34.5 \pm 1.2$ \\
9 & $33.2 \pm 2.9$ & $36.9 \pm 1.1$ & $36.9 \pm 1.0$ & $37.6 \pm 1.1$ & $37.6 \pm 1.0$ & $37.2 \pm 1.3$ & $37.3 \pm 1.3$ \\
10 & $32.3 \pm 3.1$ & $36.7 \pm 0.9$ & $36.7 \pm 1.0$ & $36.0 \pm 1.3$ & $36.1 \pm 1.2$ & $35.5 \pm 1.6$ & $36.1 \pm 1.2$ \\
\midrule
mean & $33.5 \pm 3.9$ & $38.3 \pm 4.6$ & $38.3 \pm 4.7$ & $37.7 \pm 4.2$ & $37.8 \pm 4.1$ & $37.4 \pm 4.2$ & $37.6 \pm 4.3$ \\
\midrule
vs \emph{DREME} & & & & $-0.6$ ($0.20$) & $-0.5$ ($0.25$) & $-0.9$ ($0.02$) & $-0.8$ ($0.05$) \\
vs \emph{VM-DS}$_{s=12}$ & & & & & & $-0.3$ ($0.02$) & $-0.3$ ($0.11$) \\
\bottomrule
\end{tabular}}
\end{table}

\begin{figure}[p]
    \centering
    \includegraphics[width=0.88\textwidth]{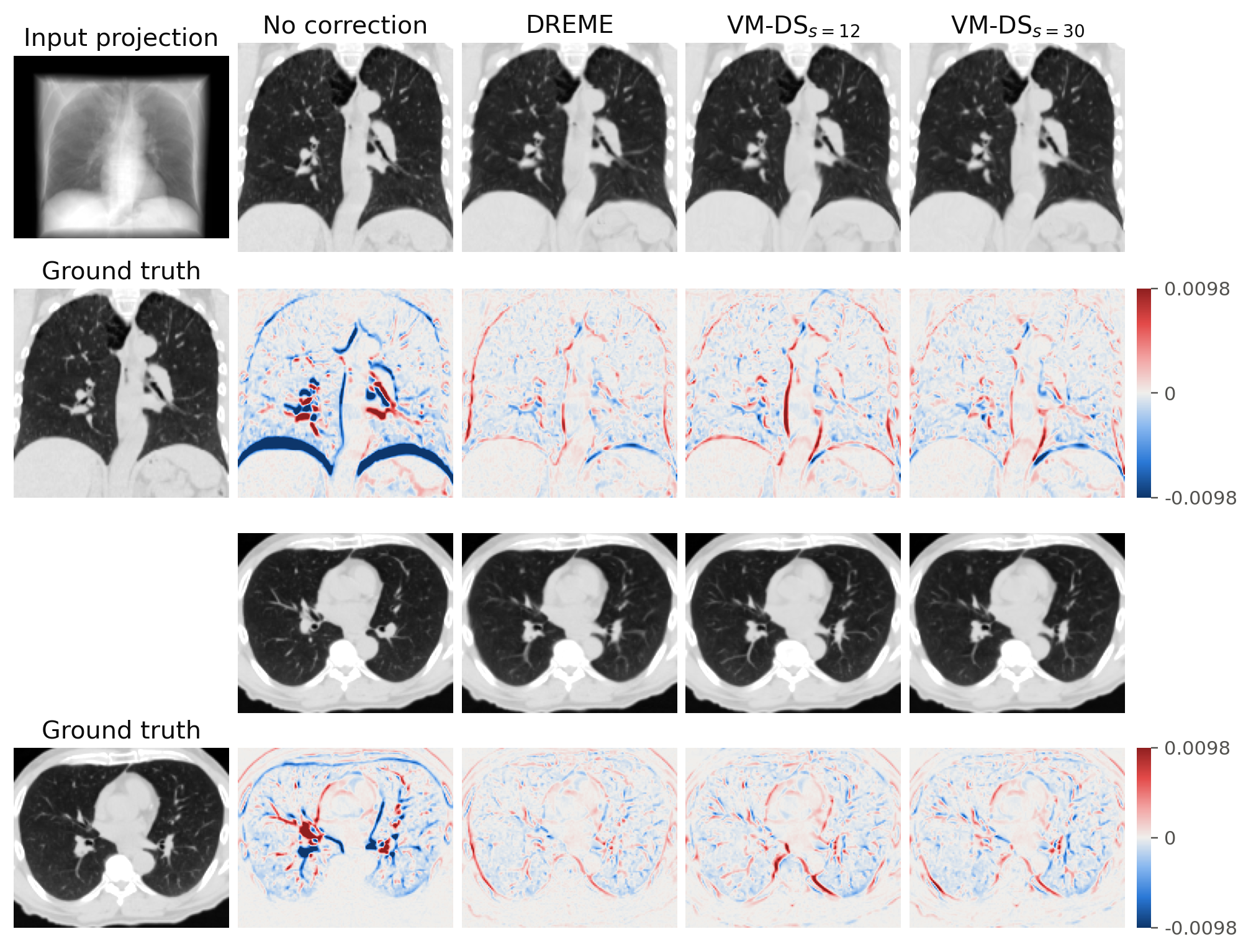}
    \caption{The held-out DIR-Lab case on which \emph{VM-DS}$_{s=12}$ was most accurate, case~$2$, at end-exhale. Upper rows show a coronal slice and lower rows an axial slice through the largest lung cross-section of the ground truth. In each pair the upper row shows the inspiration-hold reference (no correction) and the predicted volume for \emph{DREME}, \emph{VM-DS}$_{s=12}$ and \emph{VM-DS}$_{s=30}$; the lower row shows the ground truth, then the volume minus ground truth.}
    \label{fig:images}
\end{figure}

\begin{figure}[p]
    \centering
    \includegraphics[width=0.88\textwidth]{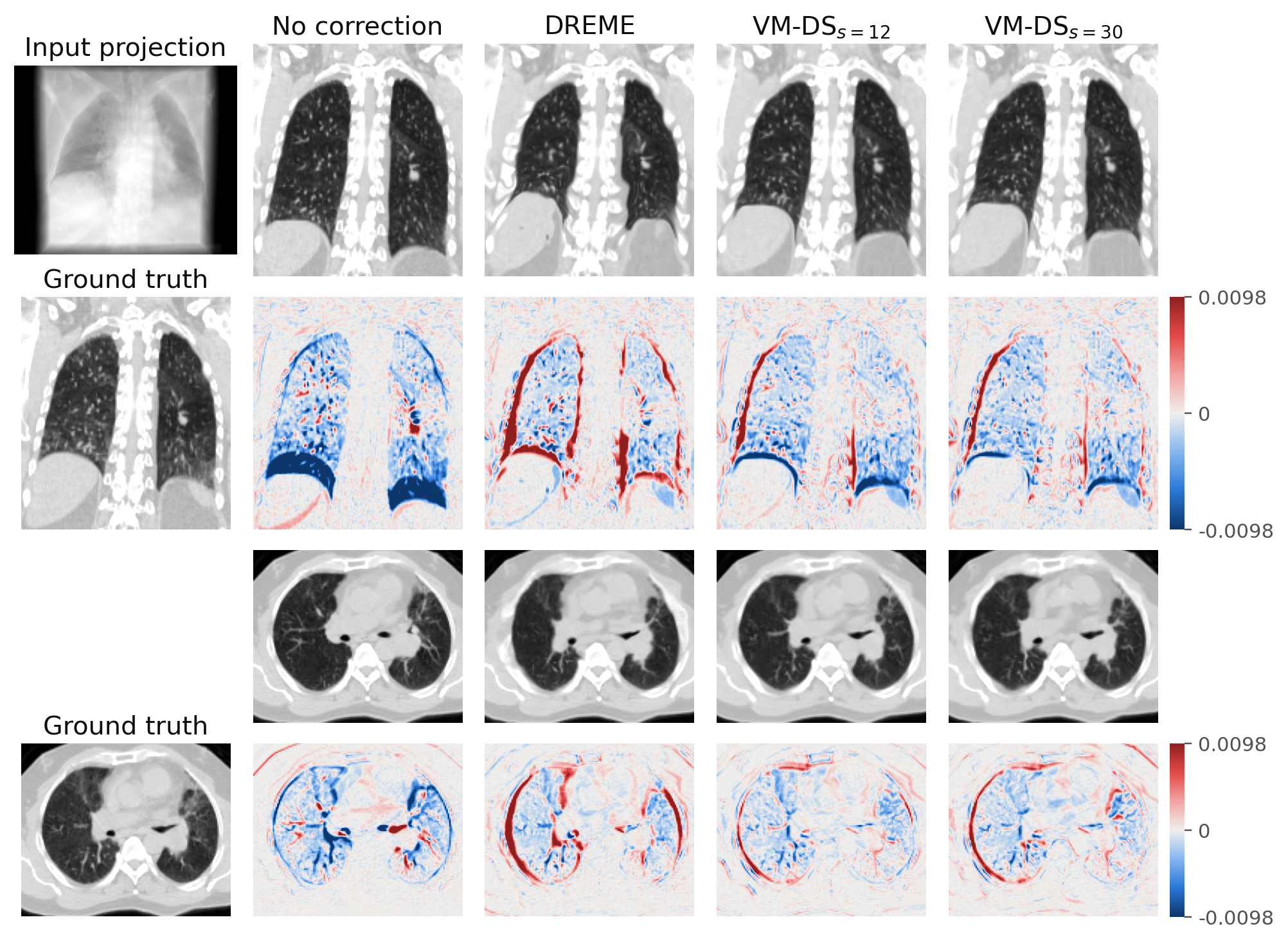}
    \caption{As Figure~\ref{fig:images}, for the held-out case on which \emph{VM-DS}$_{s=12}$ was least accurate, case~$6$.}
    \label{fig:images-worst}
\end{figure}

Tables~\ref{tab:ssim} and~\ref{tab:psnr} report image similarity on the held-out cases, on the arc and at fixed AP. Figures~\ref{fig:images} and~\ref{fig:images-worst} show the predicted volumes for the held-out cases on which \emph{VM-DS}$_{s=12}$ was most and least accurate. On the arc, \emph{DREME} gave slightly higher mean SSIM ($0.96$ against $0.95$ for both \emph{VM-DS} configurations) and PSNR ($38.3$ against $37.7$ and $37.4$~dB). The differences were not statistically significant for \emph{VM-DS}$_{s=12}$ ($p = 0.20$ for both) but were for \emph{VM-DS}$_{s=30}$ ($-0.01$ and $-0.9$~dB, $p = 0.02$ for both), which was also below \emph{VM-DS}$_{s=12}$ by less than $0.01$ in SSIM and by $0.3$~dB in PSNR ($p = 0.02$ for both). At fixed AP every method was within $0.1$~dB of its arc PSNR; \emph{VM-DS}$_{s=12}$ was again not significantly different from \emph{DREME} ($-0.5$~dB, $p = 0.25$; SSIM $p = 0.55$), and \emph{VM-DS}$_{s=30}$ was lower by $0.8$~dB ($p = 0.05$) and $0.01$ in SSIM ($p = 0.04$).

\subsection{Quantum noise}
\label{sec:res-noise}
\begin{figure}[htbp]
    \centering
    \includegraphics[width=0.8\textwidth]{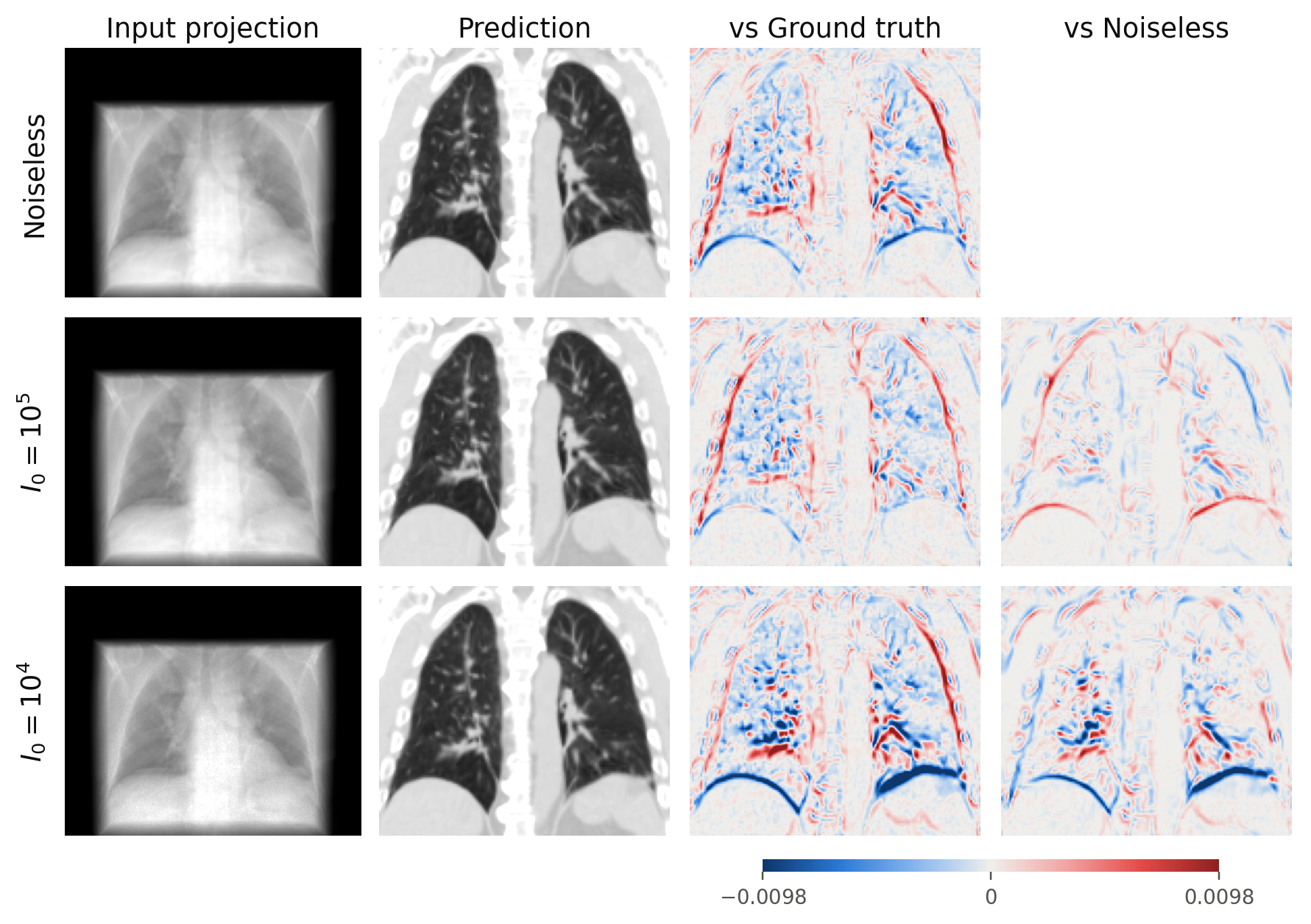}
    \caption{Effect of Poisson noise on \emph{VM-DS}$_{s=12}$ for held-out case~$4$ at end-exhale.}
    \label{fig:noise}
\end{figure}

Tables~\ref{tab:noise} and~\ref{tab:noise-ap} report held-out TRE, SSIM and PSNR for \emph{VM-DS}$_{s=12}$ under Poisson noise on the arc and at fixed AP, and Figure~\ref{fig:noise} shows its effect on one case. On the arc, mean TRE changed from $2.8$ to $2.8$~mm at $I_0 = 10^5$ ($p = 0.11$) and rose to $3.4$~mm at $I_0 = 10^4$ ($p = 0.01$), against $2.7$~mm for \emph{DREME} without noise; PSNR fell by $0.3$ and $1.6$~dB, and SSIM by less than $0.01$ and by $0.01$. At fixed AP the effect was similar: TRE was unchanged at $I_0 = 10^5$ ($p = 0.95$) and rose by $0.5$ to $3.3$~mm at $I_0 = 10^4$ ($p = 0.01$), and PSNR fell by $0.3$ and $1.4$~dB. The largest increase in TRE was on case~$6$ ($+1.1$~mm on the arc at $I_0 = 10^4$).

\begin{table}[htbp]
\centering
\caption{Held-out TRE (mm), SSIM and PSNR (dB) of \emph{VM-DS}$_{s=12}$ with Poisson noise on every projection, on the arc. Per case, mean $\pm$ standard deviation over the moving phases; the \emph{vs} row gives the paired mean difference to the noise-free model with the Wilcoxon signed-rank $p$ in parentheses.}
\label{tab:noise}
\small
\setlength{\tabcolsep}{3pt}
\resizebox{\textwidth}{!}{%
\begin{tabular}{lccccccccc}
\toprule
 & \multicolumn{3}{c}{TRE (mm)} & \multicolumn{3}{c}{SSIM} & \multicolumn{3}{c}{PSNR (dB)} \\
\cmidrule(lr){2-4}\cmidrule(lr){5-7}\cmidrule(lr){8-10}
case & noise-free & $I_0 = 10^5$ & $I_0 = 10^4$ & noise-free & $I_0 = 10^5$ & $I_0 = 10^4$ & noise-free & $I_0 = 10^5$ & $I_0 = 10^4$ \\
\midrule
2 & $1.9 \pm 0.4$ & $1.9 \pm 0.4$ & $2.4 \pm 0.5$ & $0.95 \pm 0.01$ & $0.95 \pm 0.01$ & $0.93 \pm 0.02$ & $34.4 \pm 1.4$ & $34.2 \pm 1.3$ & $31.9 \pm 1.8$ \\
3 & $2.5 \pm 0.6$ & $2.5 \pm 0.5$ & $2.9 \pm 0.8$ & $0.99 \pm 0.00$ & $0.99 \pm 0.00$ & $0.98 \pm 0.01$ & $47.7 \pm 1.9$ & $47.7 \pm 1.6$ & $46.0 \pm 1.9$ \\
4 & $2.9 \pm 0.8$ & $3.0 \pm 0.8$ & $3.6 \pm 1.1$ & $0.96 \pm 0.01$ & $0.95 \pm 0.01$ & $0.94 \pm 0.02$ & $38.1 \pm 2.1$ & $37.4 \pm 2.0$ & $36.4 \pm 1.8$ \\
5 & $3.0 \pm 0.8$ & $3.1 \pm 0.7$ & $3.5 \pm 0.9$ & $0.96 \pm 0.01$ & $0.96 \pm 0.01$ & $0.95 \pm 0.01$ & $37.4 \pm 1.2$ & $37.0 \pm 1.3$ & $36.1 \pm 1.2$ \\
6 & $3.5 \pm 0.7$ & $3.7 \pm 1.0$ & $4.7 \pm 1.3$ & $0.93 \pm 0.02$ & $0.93 \pm 0.02$ & $0.92 \pm 0.02$ & $35.6 \pm 2.3$ & $35.1 \pm 2.2$ & $34.1 \pm 1.6$ \\
7 & $3.2 \pm 0.9$ & $3.2 \pm 1.0$ & $4.1 \pm 1.1$ & $0.93 \pm 0.02$ & $0.93 \pm 0.02$ & $0.92 \pm 0.02$ & $35.0 \pm 1.8$ & $34.7 \pm 2.0$ & $33.6 \pm 1.4$ \\
9 & $2.4 \pm 0.6$ & $2.4 \pm 0.6$ & $3.0 \pm 0.6$ & $0.96 \pm 0.01$ & $0.96 \pm 0.01$ & $0.95 \pm 0.01$ & $37.6 \pm 1.1$ & $37.4 \pm 1.1$ & $35.9 \pm 1.2$ \\
10 & $2.9 \pm 0.8$ & $2.9 \pm 0.7$ & $3.2 \pm 0.7$ & $0.95 \pm 0.01$ & $0.94 \pm 0.01$ & $0.93 \pm 0.01$ & $36.0 \pm 1.3$ & $35.9 \pm 1.3$ & $35.0 \pm 1.2$ \\
\midrule
mean & $2.8 \pm 0.5$ & $2.8 \pm 0.6$ & $3.4 \pm 0.7$ & $0.95 \pm 0.02$ & $0.95 \pm 0.02$ & $0.94 \pm 0.02$ & $37.7 \pm 4.2$ & $37.4 \pm 4.3$ & $36.1 \pm 4.3$ \\
\midrule
vs noise-free & & $+0.1$ ($0.11$) & $+0.6$ ($0.01$) & & $-0.00$ ($0.02$) & $-0.01$ ($0.01$) & & $-0.3$ ($0.02$) & $-1.6$ ($0.01$) \\
\bottomrule
\end{tabular}}
\end{table}

\begin{table}[htbp]
\centering
\caption{Held-out TRE (mm), SSIM and PSNR (dB) of \emph{VM-DS}$_{s=12}$ with Poisson noise on every projection, on at fixed AP. Per case, mean $\pm$ standard deviation over the moving phases; the \emph{vs} row gives the paired mean difference to the noise-free model with the Wilcoxon signed-rank $p$ in parentheses.}
\label{tab:noise-ap}
\small
\setlength{\tabcolsep}{3pt}
\resizebox{\textwidth}{!}{%
\begin{tabular}{lccccccccc}
\toprule
 & \multicolumn{3}{c}{TRE (mm)} & \multicolumn{3}{c}{SSIM} & \multicolumn{3}{c}{PSNR (dB)} \\
\cmidrule(lr){2-4}\cmidrule(lr){5-7}\cmidrule(lr){8-10}
case & noise-free & $I_0 = 10^5$ & $I_0 = 10^4$ & noise-free & $I_0 = 10^5$ & $I_0 = 10^4$ & noise-free & $I_0 = 10^5$ & $I_0 = 10^4$ \\
\midrule
2 & $1.8 \pm 0.3$ & $1.9 \pm 0.3$ & $2.4 \pm 0.5$ & $0.95 \pm 0.01$ & $0.95 \pm 0.01$ & $0.93 \pm 0.02$ & $34.5 \pm 1.2$ & $34.1 \pm 1.2$ & $32.1 \pm 1.6$ \\
3 & $2.5 \pm 0.5$ & $2.4 \pm 0.5$ & $2.8 \pm 0.7$ & $0.99 \pm 0.00$ & $0.99 \pm 0.00$ & $0.99 \pm 0.00$ & $47.5 \pm 1.8$ & $47.7 \pm 1.6$ & $46.4 \pm 1.8$ \\
4 & $2.7 \pm 0.6$ & $2.7 \pm 0.6$ & $3.3 \pm 0.7$ & $0.96 \pm 0.01$ & $0.96 \pm 0.01$ & $0.94 \pm 0.01$ & $38.4 \pm 1.8$ & $37.9 \pm 1.8$ & $36.6 \pm 1.4$ \\
5 & $3.1 \pm 0.8$ & $3.2 \pm 0.7$ & $3.6 \pm 0.9$ & $0.96 \pm 0.01$ & $0.96 \pm 0.01$ & $0.95 \pm 0.01$ & $37.4 \pm 1.2$ & $36.9 \pm 0.9$ & $36.0 \pm 1.0$ \\
6 & $3.7 \pm 1.0$ & $3.6 \pm 0.9$ & $4.4 \pm 1.1$ & $0.94 \pm 0.02$ & $0.93 \pm 0.02$ & $0.92 \pm 0.02$ & $35.9 \pm 2.3$ & $35.5 \pm 2.0$ & $34.6 \pm 1.7$ \\
7 & $3.1 \pm 0.9$ & $3.1 \pm 0.8$ & $3.7 \pm 1.0$ & $0.94 \pm 0.01$ & $0.93 \pm 0.02$ & $0.92 \pm 0.02$ & $35.2 \pm 1.3$ & $34.9 \pm 1.5$ & $33.8 \pm 1.0$ \\
9 & $2.3 \pm 0.6$ & $2.5 \pm 0.7$ & $2.7 \pm 0.5$ & $0.96 \pm 0.01$ & $0.96 \pm 0.01$ & $0.95 \pm 0.01$ & $37.6 \pm 1.0$ & $37.3 \pm 1.2$ & $36.4 \pm 1.0$ \\
10 & $2.9 \pm 0.7$ & $2.8 \pm 0.7$ & $3.2 \pm 0.8$ & $0.95 \pm 0.01$ & $0.95 \pm 0.01$ & $0.94 \pm 0.01$ & $36.1 \pm 1.2$ & $36.2 \pm 1.3$ & $35.2 \pm 1.2$ \\
\midrule
mean & $2.8 \pm 0.6$ & $2.8 \pm 0.5$ & $3.3 \pm 0.6$ & $0.96 \pm 0.02$ & $0.95 \pm 0.02$ & $0.94 \pm 0.02$ & $37.8 \pm 4.1$ & $37.6 \pm 4.3$ & $36.4 \pm 4.3$ \\
\midrule
vs noise-free & & $+0.0$ ($0.95$) & $+0.5$ ($0.01$) & & $-0.00$ ($0.02$) & $-0.01$ ($0.01$) & & $-0.3$ ($0.04$) & $-1.4$ ($0.01$) \\
\bottomrule
\end{tabular}}
\end{table}

\section{Discussion}
\label{sec:discussion}
\begin{figure}[htbp]
    \centering
    \includegraphics[width=1.0\textwidth]{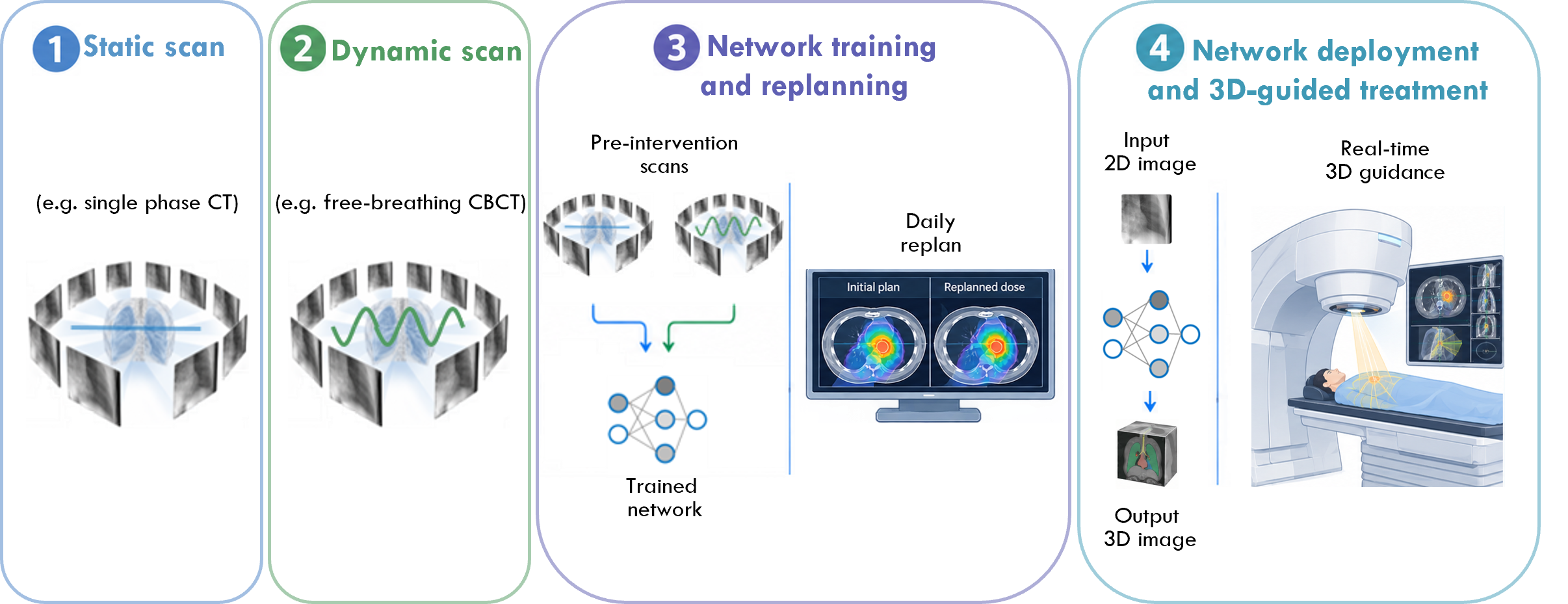}
    \caption{Our proposed workflow applied in the radiotherapy context.}
    \label{fig:rt}
\end{figure}

In this paper we introduce the \emph{VM-DS} optimization framework for volumetric imaging during image-guided interventions. The framework adapts a dual-scan regime in which an inspiration-hold CBCT acquired as standard of care to plan the navigation pathway serves as the reference, while an additional motion-inclusive CBCT, acquired while the patient is anaesthetised and ventilated, supplies the motion. Training is split into three steps (Figure~\ref{fig:fig2}). A low-rank motion model is solved against measured projections by alternating least squares, the solved model then synthesizes projections with known motion scores, and a projection encoder amortizes the fields by training on synthesized pairs alone. All learning is completed in the navigation window, and during sampling a volume is inferred from each projection in a single forward pass of $1.5$~ms. Here, while we focus on the application in image-guided biopsy, the framework could also be applied to image-guided radiotherapy (Figure~\ref{fig:rt}). In this context, the reference scan could be the planning CT or breath-hold CBCT on the day, a free-breathing CBCT is acquired to supply patient-specific motion, the daily treatment replan serves as our training window and real-time 3D motion estimation is used to guide treatment beams.

On the public DIR-Lab dataset, \emph{VM-DS} reduced the mean landmark TRE from $5.7$~mm without correction to $2.8$~mm, both on the acquisition arc and at a fixed AP angle, within a $10$-minute training window, and with statistically insignificantly performance from \emph{DREME} with its $7$-stage regime and after $17$ times longer training. Accuracy was robust to Poisson noise at $I_0 = 10^5$ and degraded by about $0.5$~mm at $I_0 = 10^4$.

The speed of the \emph{VM-DS} optimization framework leverages the existence of a known reference scan. In \emph{DREME} the INR holds $16.8$ million of the $16.9$ million learnable parameters, against $2.0 \times 10^7$ detector values in its $227$ training projections. Without the INR, Stage~1 of \emph{VM-DS} solves $71{,}361$ control points and $9$ scores for each of its $57$ projections against $5.0 \times 10^6$ detector values. With the reference fixed the model is also bilinear, so each block is a least-squares problem, solved per projection by a $9 \times 9$ Gauss--Newton step for the scores and by a few gradient steps for the control points.

The component ladder (Table~\ref{tab:ladder}) attributes the accuracy to three components. The inspiration-hold reference accounts for most of it, provided it is held fixed. Indeed, \emph{DREME}'s own curriculum improved from $11.0$ to $3.8$~mm on the high-motion development case, whereas left free to learn (\emph{DREME}$_{Iref}$) it reached $2.1$~mm on the low-motion development case against $1.4$~mm fixed. Removing stages of its curriculum on development case~$8$ traced this to stage I-b, which fits a static reference to projections acquired throughout the respiratory cycle. The solver then brings training from hours into the window, and synthesis makes the encoder usable at a fixed C-arm angle. On a single ventilated arc, each angle is seen at only one respiratory phase, so an encoder fitted to the measured projections can read the phase from the angle. Training on synthesized pairs, which place every motion state at every angle, breaks this correspondence.

As the target is delineated once, on the inspiration-hold CBCT on which the navigation pathway is already planned, and is propagated to each intra-procedural frame by the predicted DVF, no segmentation of a reconstructed volume is required during the intervention. The motion is likewise inferred from the whole projection rather than from the visibility of the target, so a nodule obscured by the mediastinum, the spine or the bronchoscope itself can in principle be localized through its correlation with visible anatomy, although occlusion was not tested here. Moreover, this may require a large enough field of view for structures, such as the diaphragm, to be persistently visible. Small, low-contrast lesions are those for which navigation is most difficult and fluoroscopic visibility poorest, so this property matters more in the interventional setting than in radiotherapy, where targets are typically larger.

In this study, a second motion-inclusive scan in the same session, against which to test, does not exist. Hence, as in \emph{DREME}, we held out projections of the training arc. On a single arc, however, held-out views neighbour training views at nearly the same angle and phase, so the standard protocol keeps angle and phase coupled and can flatter a method that learns the angle. We therefore also report every result at fixed AP, where each held-out angle is rendered at every moving phase, which is the condition of fluoroscopy at a fixed C-arm angle. The two imaging regimes agreed to within $0.1$~mm for every method, suggesting that \emph{VM-DS} does not rely on phase-angle coupling.

\begin{figure}[htbp]
    \centering
    \includegraphics[width=1\textwidth]{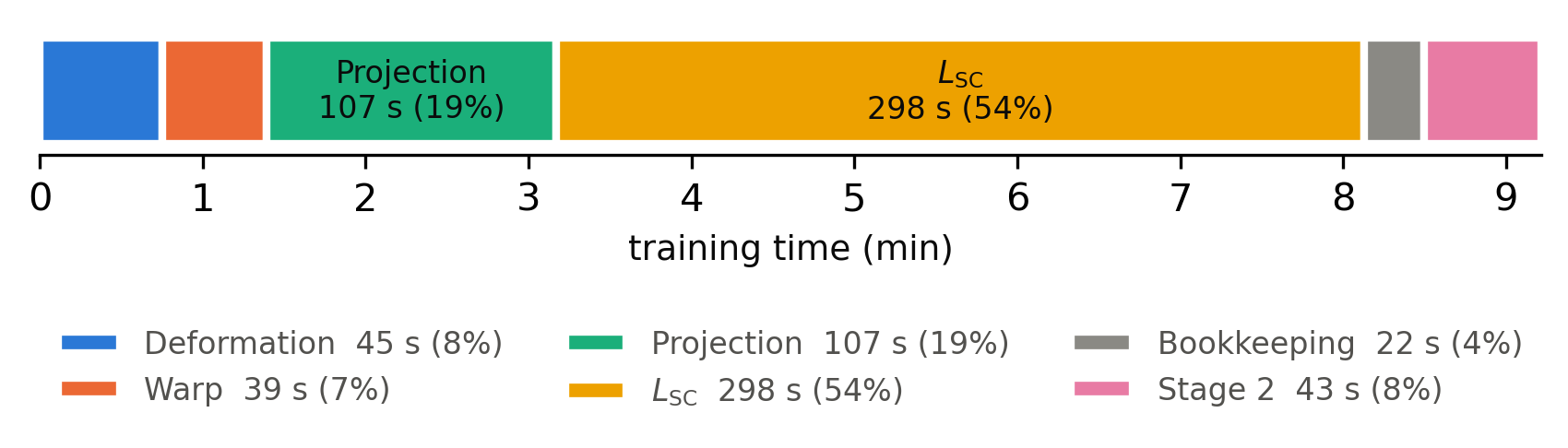}
    \caption{Training time of \emph{VM-DS}$_{s=12}$ on development case~$1$ by component. Stage~1 times are those of the run; the $L_{\mathrm{SC}}$ share is the measured difference between runs with and without it on an otherwise idle GPU, and the remainder of the score and control-point steps is divided between deformation, warp and projection in proportion to their measured forward cost per batch. Bookkeeping is the per-iteration overhead of Stage~1; Stage~2 is synthesis and encoder training, timed on an otherwise idle GPU on case~$8$. The $L_{\mathrm{SC}}$ cost of the control-point steps was measured at $K = 14$ and is scaled to $K = 23$.}
    \label{fig:lsc}
\end{figure}

Wall clock training time was the major hurdle addressed by this work and more than half of this time is currently spent on regularization (Figure~\ref{fig:lsc}). On development case~$1$ the self-consistency term $L_{\mathrm{SC}}$ accounts for an estimated $54\%$ of the training time ($5.0$ of $9.2$~minutes). Per batch of $32$ projections its inverse fixed-point iteration takes $147$~ms and its second warp $27$~ms, against $35$~ms for the deformation, $30$~ms for the warp and $83$~ms for the projection. $L_{\mathrm{SC}}$ is nonetheless load-bearing. Without it the motion field diverged on both development cases (Table~\ref{tab:ladder}, Figure~\ref{fig:ladder-curves}). A cheaper regularizer would translate directly into faster training. This forms a major direction for future study.

The held-out errors presented here should be read against the scale of the clinical task. Biopsy is recommended for lesions larger than $8$~mm \citep{Gould2013,Rivera2013}, whose radius is thus at least $4$~mm, and CBCT-informed repositioning moves the needle a mean of $4$~mm toward the lesion centre per adjustment \citep{Revelo2025}. The error of \emph{VM-DS} is smaller than the corrections currently made on the basis of repeat spins, but at $2.8$~mm there is still room for improvement. Inference at $1.5$~ms per projection leaves ample margin for real-time use during fluoroscopy.

In terms of limitations, all acquisitions were simulated from 4DCT phase volumes, via forward-projection without scatter or beam hardening. We previously evaluated the use of diaphragm tracking to achieve 4D imaging using images acquired on a C-arm interventional system for a deformable phantom \citep{Reynolds2024}. This setup could be useful for future work evaluating \emph{VM-DS}.

Real inspiration-hold imaging reaches a deeper inspiration than the end-inhale phase used as its proxy, so the simulated deformations potentially understate the motion that could be encountered in real clinical settings. Additionally, respiration was modelled as a triangular cycle through ten discrete phase volumes, without hysteresis or cycle-to-cycle variation. The regime assumes general anaesthesia with machine-paced ventilation, as in bronchoscopic biopsy; percutaneous biopsy under sedation falls outside it. Fixed-AP accuracy rests on $49$ nearly identical views at each of five moving phases, so each case contributes five motion states at that angle.

Each held-out case was run once, and eight held-out cases from a single dataset are a modest sample. However, the DIR-Lab data used here are unique in being both pubicly available, evaluated on many published methods, while supplying clinician-delineated landmarks for evaluation. We are currently working towards ethics approval for a phase 1 data collection clinical trial in the interventional setting.

The motion model inherits the limitations of a B-spline decomposition. This low-rank representation assumes that motion is a superposition of three components, sufficient for respiration \citep{Li2011PCA} but not for coughing or cardiac motion, which the ventilated, anaesthetised patient largely excludes. The current representation assumes smooth motion, so the sliding of the lung against the chest wall is not captured. Additionally, the inspiration-hold reference is assumed to share the anatomy of the ventilated scan, so changes between them, such as atelectasis developing under anaesthesia, are not modelled.

\section{Conclusion}
\label{sec:conclusion}
Cutting-edge bronchoscopic techniques still yield false negatives in $20\%$ of cases and fail to yield clear diagnostic decisions in $30\%$. Intraprocedural 3D imaging drastically improves results, but each 3D spin represents a static snapshot of an anatomy that is continuously changing. Here we propose the \emph{VM-DS} optimization framework that trains within a $10$ minute navigation window and enables real-time volumetric imaging from a single x-ray projection during sampling. This framework is generalizable and may find application in radiotherapy and other image-guided interventions.

\section*{Acknowledgements}
This work was funded by the NSW Government through the Cancer Institute NSW and by the Australian Government by the National Health and Medical Research Council.

\section*{Code and data availability}
The code is available at https://github.com/Image-X-Institute/Voxelmap/tree/VM-DS. The DIR-Lab 4DCT data and landmarks are publicly available from the DIR-Lab website: https://med.emory.edu/departments/radiation-oncology/research-laboratories/deformable-image-registration/downloads-and-reference-data/4dct.html.

\section*{Competing interests}
The author has filed a patent relating to the method disclosed in this paper.

\bibliographystyle{plainnat}
\bibliography{introduction}

\end{document}